# Large-Area Atomically Flat Monocrystalline Gold Flakes: Recent Advances, Applications, and Future Potential

*Amro O. Sweedan* *[1,2], Kefan Zhang [2], Muhammad Y. Bashouti [1,2] and Thorsten Feichtner* *[3]*

[1] *The Ilse-Katz Institute for Nanoscale Science & Technology, Ben-Gurion University of the Negev, POB 653, Beer-Sheba Campus, Building 51, 8410501, Israel.*

[2] *Department of Solar Energy and Environmental Physics, Swiss Institute for Dryland Environmental and Energy Research, J. Blaustein Institutes for Desert Research, Ben-Gurion University of the Negev, Midreshet Ben-Gurion, Building 26, 8499000, Israel.*

[3] *Nano-Optics and Biophotonics Group, Experimental Physics 5, Institute of Physics, Am Hubland, University of Würzburg, Germany.*

Keywords

Monocrystalline gold, gold flakes, gold platelets, wet-chemical synthesis, plasmonics, nano photonics, sensing, gold structuring

## Abstract

High aspect ratio oblate polygonal gold crystals - such as triangular and hexagonal platelets - have attracted considerable interest due to their extraordinary physical, chemical, and mechanical properties. Commonly referred to as "gold flakes," these structures exhibit atomically flat surfaces, µm² areas with nanometric thickness, and a monocrystalline morphology. Since their first discovery by John Turkevich in 1951, considerable progress has been made in shape-controlled synthesis and large-scale production, unlocking steadily new opportunities for ever more advanced applications. This review explores large-area gold flakes with lateral dimensions spanning from hundreds of nanometers to millimeters, emphasizing their unique properties. We provide a comprehensive overview of key developments, from early discoveries, synthesis approaches, and fabrication techniques to recent breakthroughs. Emphasis is placed on the integration of gold flake as functional building blocks in photonics (e.g., for nanoantennas), sensing, nanoelectronics, biomedicine, and beyond. We conclude with a discussion of emerging roles and future developments of this unique class of materials.

## Introduction

Gold nanostructures, particularly wet-chemically synthesized nanoparticles and large area nanofilms with a confined thickness, have garnered significant interest due to their unique physical, chemical, and optical properties [1-4]. These materials have been widely utilized across diverse fields, including photonics [5-8], electronics [9-12],  and diagnostics [13-15], among others [16-20]. Various possible morphologies of low-dimensional gold have been reported, including spherical nanoparticles [18, 21], nanorods [22, 23], nanoflakes [24, 25], and other shapes [26-29]. Among these, a particularly notable class consists of high-aspect-ratio oblate nanostructures, typically triangular or hexagonal platelets, commonly and within this review called "gold flakes", but also "gold platelets", "gold prisms" or "gold sheets" [30] (**Figure 1**). These particles are extraordinary due to their large area atomically flat surfaces, monocrystalline nature, and therefore boundary-free morphology, making them excellent candidates for applications requiring precise machinability, large conductivity, mechanical stability, or a combination of these[10, 31]. To this day, high-yield synthesis with moderate to excellent control over the particle morphology has been well established [32, 33]. In particular, for flakes, a decent tuning of thickness and lateral dimensions has been demonstrated [6, 34].

This bottom-up approach to realize flat, thin metal films differs from conventional fabrication techniques, such as gold evaporation and subsequent lithography, milling, polishing, thinning or template stripping, which always yield polycrystalline thin films, showing grain boundaries and structural defects that degrade material quality and performance [8, 35, 36].

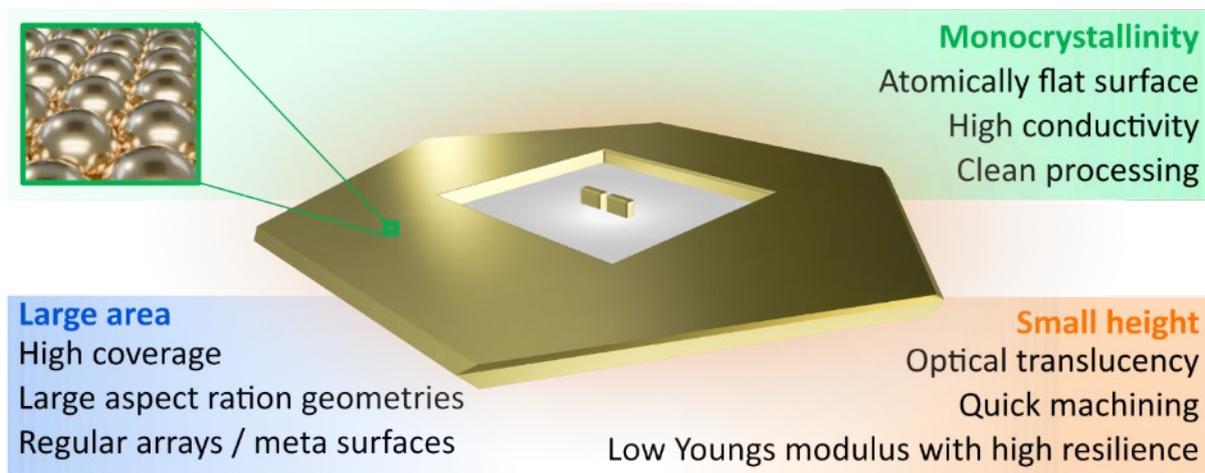

**Figure 1: Gold flake properties overview**. The three main properties of large aspect ratio oblate Au monocrystals and the main resulting benefits for their application.

The earliest documented observation of gold flakes dates back to 1951, when John Turkevich and colleagues identified them as by-products in complex reaction mixtures during electron microscopy studies [37]. Over the following decades, research focused on understanding the formation mechanism of these structures, optimizing the synthetic protocols to reproducibly achieve large-area and scalable production, and obtaining control on shape formation for unlocking the full potential of the flakes as universal building blocks for nano-devices [8, 38-42]. To the best of our knowledge, the first high-yield, shape-controlled synthesis of noble metal flakes was reported in the early 2000s by Jin *et al.* [43] using silver, paving the way for subsequent efforts to realize large-area monocrystalline gold flakes. Independent research groups later introduced various high-yield methods for establishing controlled growth of gold flakes exceeding a diameter of a couple of hundred nanometers [44-46], further advancing their scalability and practical applications. Today, gold flakes can be synthesized using a range of different strategies, mostly variations of the classical wet-chemical recipes [8, 10, 47], but also e.g., physical methods [48, 49].

While extensive research on small-area gold nanostructures can be found [30, 50-52], studies on large-area gold flakes—ranging from micrometer to millimeter scale while still showing nanometric thickness confinement—remain relatively limited until today. However, their increasing utilization, whether employed as-synthesized in a classical bottom-up fashion or further structured using top-down techniques such as focused ion beam (FIB) milling, will benefit from a comprehensive collection of the existing knowledge as a starting point for the next decades of gold flake-based research. This review summarizes large-area gold flake structural and physical properties, synthesis techniques, applications in photonics, (bio)sensing and nanoelectronics, as well as their potential in next-generation technologies. Therefore, some applications are highlighted and described in detail, while the exhaustive overview is captured within concise tables to make the overview as complete as possible.

## Gold flake structural properties and growth mechanism

Gold flakes —along with their less prevalent geometrical variations, such as nanosheets and nano disks [8, 10, 47], are characterized as oblate crystal with a C3 or C6 symmetry along its short axis, their shape being triangular, hexagonal or anything in-between (see e.g. **Figure 2** (A) or (D)). Geometrically, these crystals are defined by two parallel polygonal surfaces, with lateral dimensions ranging between a few nanometers and millimeters [36, 53]. Owing to their dimensions, which laterally often exceed several hundred nanometers and vertically rarely go below 20 nm, the bare flakes typically do not exhibit phenomena associated with nanoscale confinement and can be described entirely using classical models [2, 54-56]. The crystal structure of solution-prepared gold flakes is typically monocrystalline, exhibiting a face-centered cubic (fcc) lattice [30, 45, 57-59], with a boundary-free configuration along their lateral plane [35, 60]. The upper/lower facets are composed of nearly atomically {111} crystal faces [57-59, 61]. Edges commonly expose {100} or {110} planes, high-index facets, or even {111}, depending on the flake's thickness and specific growth conditions [1, 61, 62] (see **Figure 2** (A)-(C)). High-resolution transmission electron microscopy (HRTEM) and convergent beam electron diffraction (CBED) analysis both show that gold flakes contain at least one twin plane parallel to their {111} upper surface (**Figure 2** (D)). This twin boundary is a result of an initial atomic-scale stacking fault, where the fcc lattice inversion creates a symmetrical plane of reflection across the flake plane [57, 59, 61, 63] (**Figure 2** (F)-(G)). This is one origin of the enhanced in-plane growth of the flakes during synthesis (a simulation respecting this principle can be found here [64]).

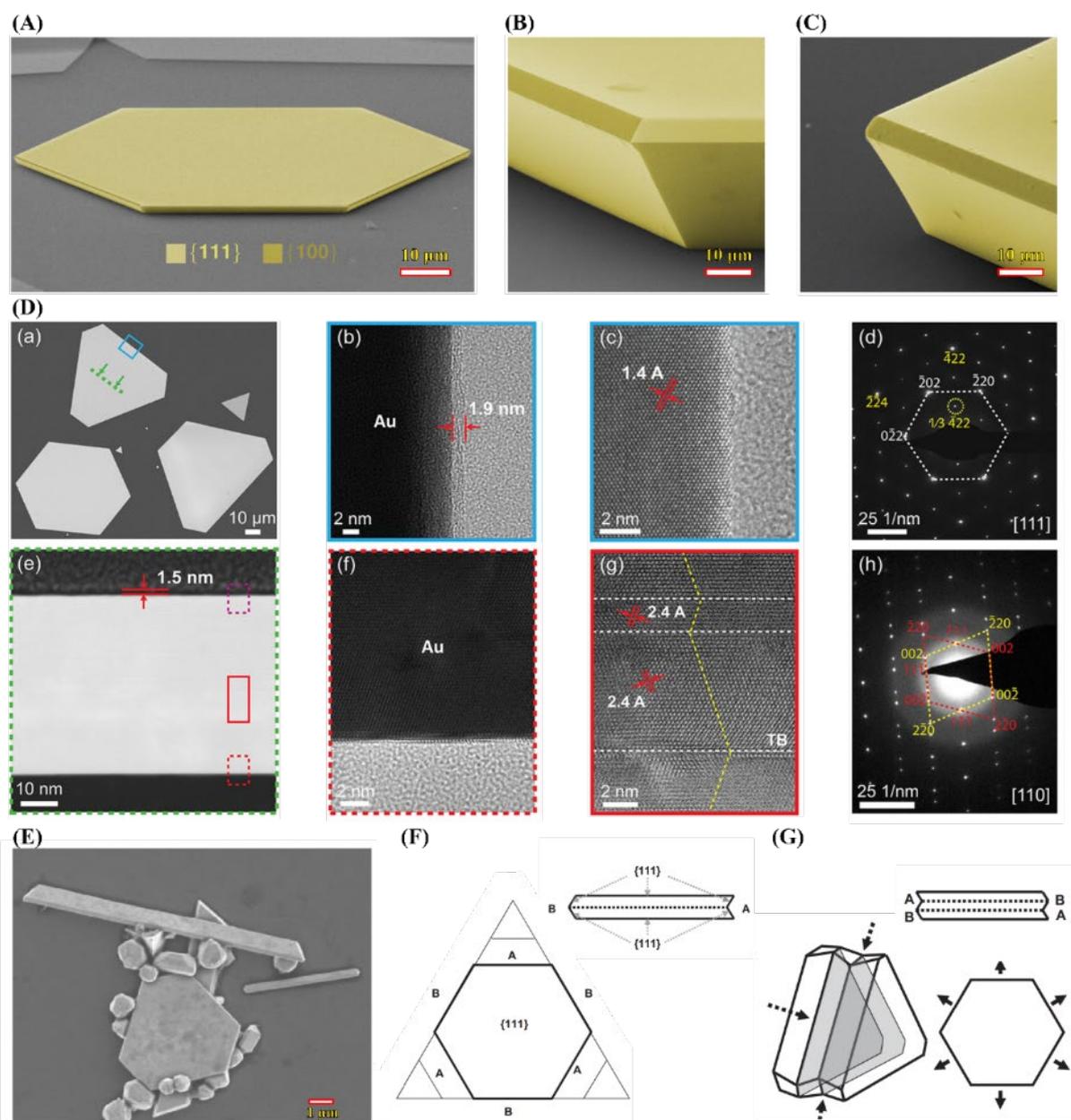

Figure 2: **Structural properties of gold flakes.** (A) SEM image of an Au monocrystalline flake tilted at a 75° angle. (B-C) Close-up high-resolution SEM images of two corners of the flake, also tilted at 75°. Artificial coloration highlights different crystallographic planes on the facets: light yellow for {111} and dark yellow for {100}. Panels (A)-(C) adapted from Boroviks *et al.* [65]. Copyright © 2018 Optical Society of America. (D) Microstructural and crystallographic properties of gold flakes. ($D_a$) Top-view SEM image of gold flakes. ($D_b$) Planar-view bright-field TEM image of the edge of a solution-grown gold flake, showing a 1.9 nm thick organic layer on the side facet. ($D_c$), ($D_d$) Atomic-resolved HRTEM image and selected area electron diffraction pattern (SAED) of the same region. ($D_e$) Cross-sectional STEM image of a flake after FIB cross-sectioning. ($D_f$) Atomic-resolved HRTEM image of the flake–glass substrate interface region. ($D_g$) Higher magnification HRTEM image and ($D_h$) SAED pattern of the twinned region of the flake, with 2.4 Å spacing between the {111} planes of fcc-Au, twin boundaries, and mirrored diffraction spots. Panel D Adapted from Kiani *et al.* [57]. Copyright © 2022 The Authors. (E) Diversity of shapes synthesized in a homogeneous reaction environment, including rods, tapes, flakes, tetrahedra, and isotropic particles. (F) Sketch of a single twin plane Au flake, where alternating sides display A-type and B-type faces. The reentrant grooves on the A-type faces promote rapid growth, which is halted when the face grows out, leaving behind a triangular prism with slower-growing B-type faces. (G) Sketch with two parallel twin planes. All six sides display A-type faces (denoted by dashed arrows) with reentrant grooves. This enables each A-type face to regenerate those adjacent to it, facilitating rapid growth in two dimensions (solid arrows). Panels (E)-(G) reprinted with permission from Lofton *et al.* [61]. Copyright © 2005 WILEY-VCH Verlag GmbH & Co. KGaA, Weinheim.

According to LaMer's nucleation model, after the reduction of gold ions, the resulting gold atoms reach a supersaturation threshold before spontaneously forming small clusters, which then evolve into stable nuclei [66]. Surface energy considerations result in favor of {111} facets, leading to the formation of symmetrical seeds, as these configurations minimize free energy [67]. Anisotropic shapes will be obtained only when this rule is broken. This can be achieved by employing capping agents that selectively regulate growth rates across different facets while gold ions are being reduced in a kinetically controlled pathway [32, 42, 68]. Experimentally, an efficient capping agent changes the order of free energies for different crystallographic facets of gold nanocrystals through selective chemisorption [60, 68]. These include surfactants, small-molecules, atomic adsorbates and biomolecules with molecular recognition capabilities [32, 69]. Alternatively, a template can impose external geometric constraints, accelerating seed formation and guiding crystal growth along specific directions [70], leading to anisotropic morphologies even in the absence of capping agents. Specifically, rapid nucleation induced by the template's constraints can lead to the formation of twinned seeds, where internal structural defects naturally promote anisotropic growth without the need for additional shape-directing agents [57, 70].

An example of a capping agent is polyvinylpyrrolidone (PVP), a strongly binding surfactant preferentially onto Au {111} facets, suppressing a crystals vertical growth while promoting lateral expansion [71]. Similarly, halide ions (I⁻, Br⁻, Cl⁻) exhibit a high affinity for {111} surfaces, directing nanoplate growth while also functioning as etchants that eliminate undesired nucleation sites [57, 72, 73]. The final morphology of gold flakes is therefore influenced by reaction parameters such as precursor concentration, reduction kinetics, and temperature [30]. A slow reduction rate after the nucleation phase favors the controlled deposition of gold atoms onto pre-existing seeds, preventing uncontrolled aggregation into new seeds and therefore promoting the formation of large-area flakes [32, 74]. Additionally, lateral diffusion rates on {111} facets exceed vertical stacking rates under the influence of surfactants, further enhancing the formation of 2D nanostructures [61]. The self-assembly of small nanoflakes into larger crystalline domains has also been observed, driven by high surface energy along lateral facets [75, 76].

The formation of multiple shapes, such as triangular or hexagon flakes in a single homogeneous reaction, as well as their occurrence across different synthesis methods  (see **Figure 2** (E)) was explained by Lofton *et al*. [76] and attributed to the formation of twin plane defects forming during initial nucleation, which yield both concave (A-type) and convex (B-type) surface features along the nanoplate edges (**Figure 2** (F)-(G)). The interplay between the faceting tendency between the two types dictate the final morphology. With this understanding, researchers have successfully synthesized large-area monocrystalline gold flakes with geometric control for further applications [6, 32-34]. In addition to the coexistence of multiple shapes within a single reaction, Großmann *et al*. [77] reported stepped thickness variations occurring within individual flakes.

## Synthesis routes and manipulation of large-area gold flakes

In the last two decades various synthesis approaches have been developed to realize gold flakes. Despite their diversity, all methods follow a common principle: promoting lateral growth while restricting vertical expansion to realize a thickness confinement [62, 68, 78, 79].

Gold flakes synthesis typically relies on the reduction of gold precursors, where gold ions ($Au^{3+}$) are converted into their neutral metallic state [10, 47]. The involved ions can be performed through classical wet-chemical routes [8, 10, 35, 47], physical methods [48, 49], bioinspired techniques [45, 80], and hybrid approaches that combine multiple methodologies [81, 82].

Synthesizing large-area gold flakes requires a precise balance between thermodynamic stability and kinetic control. Under purely thermodynamic constraints the formation of isotropic (fcc) structures is favored, resulting in compact geometries grown homogeneously in all directions [68]. To instead achieve two-dimensional (2D) morphologies, a clear separation of nucleation and subsequent growth conditions is required [29]. Control over the synthesis process begins with the reduction of gold precursor ions (e.g., $AuCl_4^-$) to zerovalent gold atoms through high initial concentration and/or temperature, which then aggregate into crystalline seeds [64] — a process primarily governed by thermodynamic conditions, particularly surface energy minimization [68, 83]. Under appropriate conditions, a few stacking faults are introduced in these seeds and the subsequent growth along certain crystallographic directions is suppressed. Under thermodynamically dominated conditions at lower temperatures, growth preferentially occurs at low-energy facets, leading to lateral expansion rather than isotropic (spherical) morphology [68, 84].

Growth can proceed in a template-free colloidal environment [60, 85] or through templated approaches using rigid structures, such as ITO [86] and graphene [87], or soft structures, including polymers [88], and even protein fibers [42]. Additionally, capping agents that selectively adsorb onto basal planes, along with stabilizers and other interfacial components, play a crucial role in controlling morphology and preventing aggregation [30, 68, 89]. Selected examples of flake synthesis approaches are presented in **Figure 3** and **Figure 4.** For instance, Farkas *et al.* [90] employed a classical wet-chemical reduction method within a gel matrix, creating a gradient of reducing agents across the medium. This approach yielded flakes of varying sizes, spatially organized within the gel (**Figure 3** (A)). Lv *et al.* [53] and Krauss *et al. (now Schatz et al.)* [70] adopted template-directed synthesis routes, with the later utilizing rigid glass surfaces as templates (**Figure 3** (B)), whereas the former demonstrated a bio-inspired approach using filamentous proteins that served as both a structural template and a reducing agent (**Figure 3** (C)). The resulting flakes reach macroscopic size, up to millimeters in lateral dimensions with thicknesses ranging from 0.4 - 1.3 µm, visible to the naked eye [42].

In wet chemical synthesis, there appears to be a lower limit to the aspect ratio (i.e., the ratio of thickness to lateral area). For large-area flakes, the minimal achievable thickness is typically around 10 nm. If this thickness is still too great for a particular application, it can be further reduced through thinning or etching procedures [36] (**Figure 3**(D)).

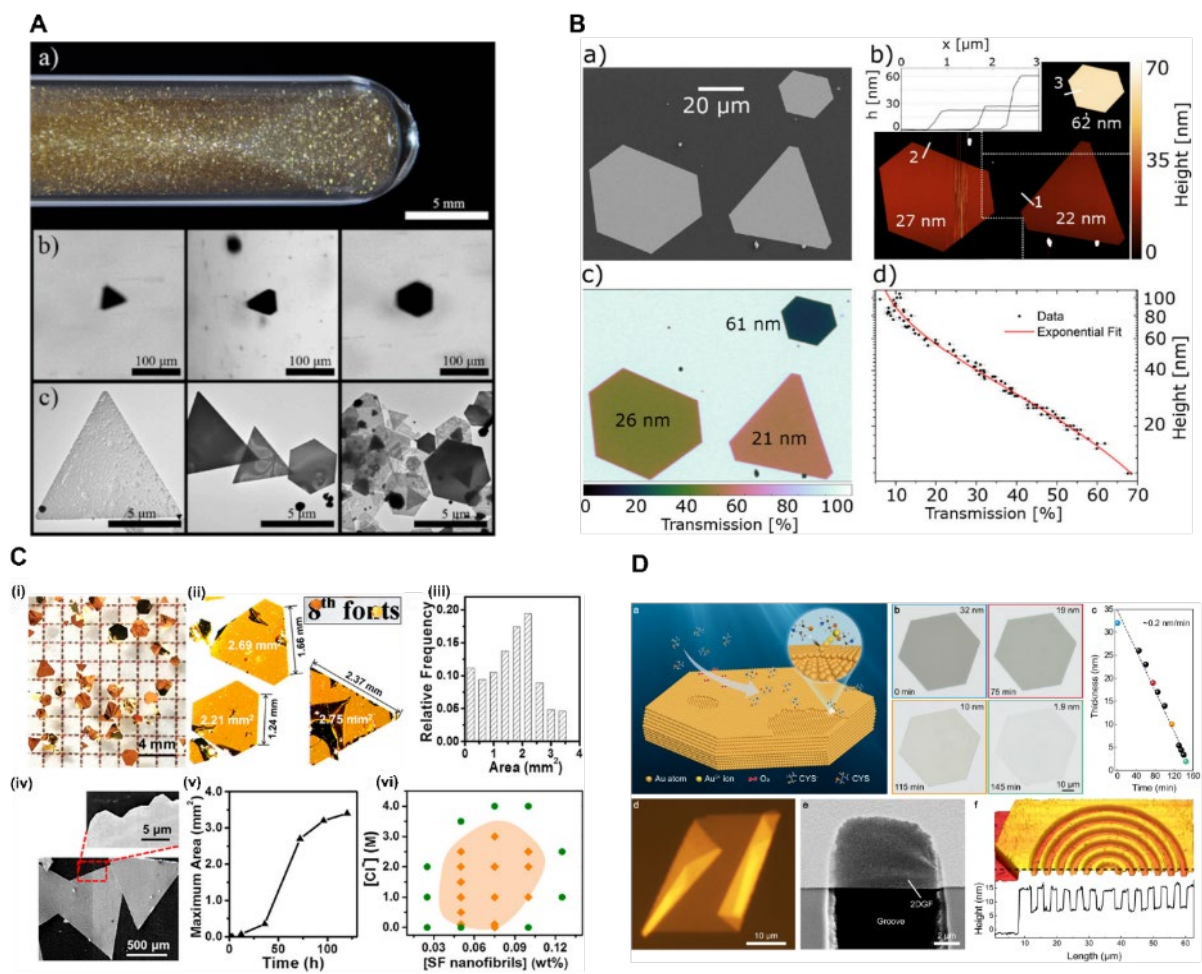

Figure 3: **Gold flake synthesis approaches.** (A) Matrix-based (template-based) chemical synthesis of gold flakes. (A$_a$) Optical photograph of the gold flakes inside the matrix; the appearance of bulk gold optical property is visible as the flake size grows from left to right. (A$_b$) Optical images of synthesized flakes. (A$_c$) TEM images of extracted flakes. Adapted from Farkas *et al.* [90] .Copyright © 2021 The Authors. (B) On substrate synthesized large-area gold flakes. (B$_a$)-(B$_c$) SEM, AFM, and optical transmission images of the identical individual gold flakes. (B$_d$) Optical transmission as a function of flake thickness. Reprinted (adapted) with permission from Krauss *et al.* [70]. Copyright © 2018 American Chemical Society. (C) Millimeter scale gold flake synthesized biochemically using nanofibrils. (C$_i$), (C$_{ii}$) Optical images of millimeter-sized gold flakes alongside a size comparison with 8-point Roman font. (C$_{iii}$) Histogram showing planar area distribution. (C$_{iv}$) SEM images of gold flakes during synthesis, the zoom showing a growth region. (C$_v$) flake area against growth time, (C$_{vi}$) Concentration window of SF nanofibrils and Cl⁻ to synthesize 2D Au crystals; Au crystal surface area > 1.0 mm$^2$ = orange; area < 1.0 mm$^2$ = green. Reprinted (adapted)  with permission from Lv *et al.* [53]. Copyright © 2018 American Chemical Society. (D) Generation of ultra-thin (single-digit nanometer) gold flakes via chemical etching. (D$_a$) Schematic of the etching process. (D$_b$) Optical transmission micrographs at different etching times. (D$_c$) Thickness of the gold flake as a function of etching time. (D$_d$) Optical reflection micrograph showing a folded gold flake. (D$_e$) SEM image of a gold flake suspended over a groove. (D$_f$) AFM image of a locally etched gold flake patterned into concentric rings. Adapted from Pan *et al.* [36]. Copyright © 2024, The Author(s).

An alternative way to categorize gold flake synthesis is based on procedural frameworks, such as single-step seedless approaches [8, 70], multi-step seed-mediated synthesis [58, 91-93], or distinctions between template-directed [42, 44, 57] and template-free strategies [50], surfactant-assisted methods [34, 94], polymer-assisted synthesis [81, 90], liquid crystal-based approaches [75, 95], and on-substrate fabrication techniques [57, 86, 96], among other classifications and variations [50].

Among chemical synthesis techniques, polyol-based wet-chemical synthesis [1, 97, 98] and its modifications have emerged as the most widely employed methods for producing gold flakes [30, 57, 70, 76, 79, 86]. In this method, a gold precursor is reduced in a polyol medium—typically ethylene glycol—which serves as both the solvent and weak reducing agent. Alternatively, external reducing agents such as citrate or aniline may be added [60, 99]. The final flake morphology is governed by reaction parameters, including precursor concentration, temperature, reaction time, pH, and the presence of surfactants or capping agents [30, 100]. Commonly used additives such as polyvinylpyrrolidone (PVP) and cetyltrimethylammonium bromide (CTAB) serve as stabilizers or surfactants [76, 101]. They are believed to selectively bind to specific crystallographic facets, directing anisotropic growth while maintaining colloidal stability and preventing aggregation [1, 32].

Beyond PVP and CTAB, a diverse range of molecular species—including other surfactants [100, 101], polymers [44, 101], biomolecules [42, 102], small organic molecules [103, 104], adsorbed gases [82, 105], and even atomic species such as metal ions [6, 57, 73]—have been explored to fine-tune flake growth. The careful selection and combination of these components enable precise control over size and shape evolution [68, 73, 89, 106]. For instance, Qin *et al.* [34] demonstrated the tunability of flake thickness by adjusting the ratio between gold precursor ions and the directing surfactant that reduces them (**Figure 4(A)).** These additives can be introduced during the initiation of the metal reduction or continuously throughout the reaction [58, 91-93], underscoring the complex interplay between reaction components and their concentration in space and time in defining flake morphology.

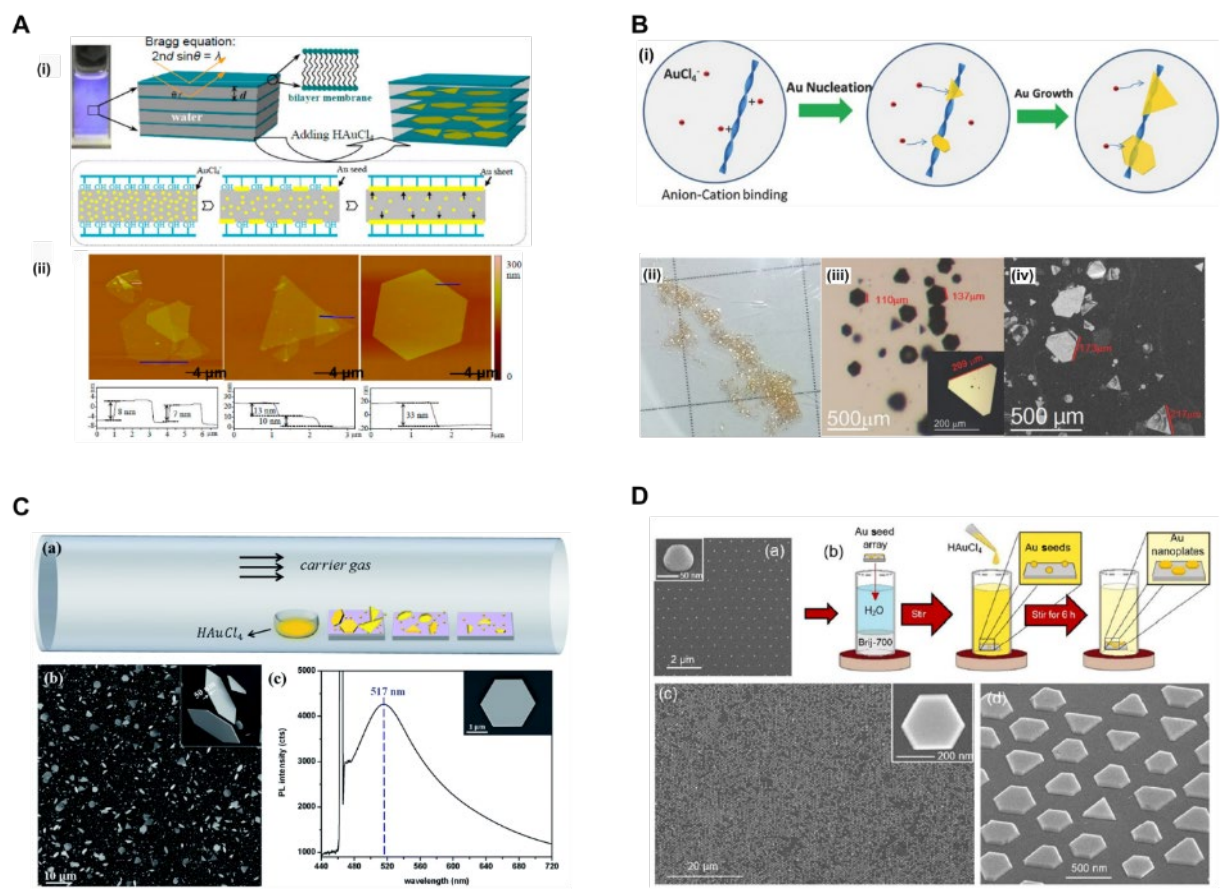

Figure 4: Large-area gold flake synthesis — advanced strategies. (A) Thickness-controlled directed synthesis of gold flakes. (A$_i$) Schematic procedure describing the water layers sandwiched directed synthesis of gold flake by lamellar bilayer membranes of a self-assembled nonionic surfactant. (A$_{ii}$) AFM scans of gold flakes obtained at different precursor concentrations and corresponding height profile curves. Reprinted (adapted) with permission from Qin *et al.* [34]. Copyright © 2013, American Chemical Society. (B) Bioinspired synthesis of macroscopic monocrystalline gold flakes using amyloid fibrils as templates. (B$_i$) Schematic illustration of the synthesis procedure, (A$_{ii}$) visual observation of resulting flakes on a grid (0.9 × 0.9 cm²), (B$_{iii}$) optical microscopy images, and (A$_{iv}$) scanning electron microscopy (SEM) image highlighting the structural features. Panel B reprinted with permission from Zhou *et al.* [42]. Copyright © 2015 WILEY-VCH Verlag GmbH & Co. KGaA, Weinheim. (C) Chemical vapor deposition-based gold flake growth. (C$_a$) Schematic of the experimental setup. (C$_b$) SEM image of the resultant gold microplates. (C$_c$) PL spectrum of a single gold microplate. Panel C Reproduced from Wang *et al.* [105] with permission from the Royal Society of Chemistry. (D) Combined top-down and subsequent bottom-up approach for gold flake arrays synthesis. (D$_a$) SEM image of a periodic array of gold nanostructures acting as seeds for flake growth. (D$_b$) Schematic representation of the solution-based growth mode. (D$_c$)-(D$_d$) SEM images of the resulting flake array. Reprinted with permission from Demille *et al.* [81]. Copyright © 2021, Tsinghua University Press and Springer-Verlag GmbH Germany, part of Springer Nature.

Biological synthesis has emerged as a promising gold flake synthesis method, leveraging biomolecules and molecular biological elements such as microorganisms, viruses, plants, proteins, and DNA molecules as templates, stabilizers, or reducing agents [69]. The first biosynthesized gold flakes exceeding 0.5 μm were reported by Shankar and colleagues [45, 107], paving the way for subsequent studies that refined the synthesis process using various biomolecules. These include amino acids [103, 104], plant-derived molecules [45, 108], polysaccharides [109, 110], and fungal-based extracts [111].

With later technological applications in mind, the biological or so-called "green" methods offer an environmentally friendly, biodegradable, and biocompatible synthesis alternative avoiding residues of toxic synthesis components in the final gold flake ensemble [69]. These sustainable approaches eliminate the need for organic solvents, harsh chemicals, stabilizers, conventional surfactants, and toxic polymers or crosslinkers, replacing them with benign biomaterials, instead, significantly reducing environmental toxicity and biological hazards. For example, Zhou *et al.* [42] demonstrated the synthesis of macroscopic gold flakes under mild, green conditions by utilizing amyloid fibril proteins, which serve multifunctional roles as reducing, directing, and stabilizing agents (**Figure 4 (B)**).

Physical flake synthesis approaches are less common. They incorporate physical tools to drive the reduction process by supplying the necessary energy and/or electrons. This includes vapor techniques [105, 112], typically employing thermolysis and a carrier gas system (**Figure 4 (C)**), photoreduction, where photons facilitate electron transfer [49], electroreduction, which utilizes an electrode as an electron source [113], as well as microwave-assisted synthesis [114]. Finally, there are approaches combining multiple techniques, including classical fabrication methods. For example, predetermining the flake positions by nano imprint lithography and subsequent seed placement by a sophisticated single crystallite synthesis yielded arrays of flakes (**Figure 4 (D)**).

A more versatile strategy for deterministic placement of a single gold flake relies on synthesizing an ensemble of flakes with varying diameters and thicknesses on a well-defined substrate. From this ensemble, the flake exhibiting the desired geometrical and optical properties is selected and subsequently transferred to the target location using a polymer-assisted transfer technique [70, 115], a method commonly handy for rapid prototyping.

## Applications

Gold flakes exhibit distinct chemical and physical properties that make them advantageous for a wide range of applications [30]. However, in cases where conventional synthesis of simple

geometries—such as triangular, spherical, or hexagonal particles—results in limited surface area and restricted geometric flexibility [116-118], large-area flakes become essential [119-124]. In such contexts, the ability to structure a material laterally with nanometer-scale precision and strictly controlled thickness—while maintaining large-area single crystallinity—is critical. These attributes are particularly important in nanophotonics and plasmonics, where gold flakes serve both as versatile substrates for top-down fabrication of complex nanostructures and as functional components in standalone devices.

The physicochemical properties of gold flakes are intrinsically linked to their structure and can be precisely tuned to meet specific application requirements. Key parameters that can be modified include edge length, thickness, structural design, molecular functionalization, and tip morphology [2, 30, 55, 125, 126]. For instance, the characteristic red color of nanometer-sized gold crystals [127] gradually fades as the material transitions to the macroscale, where large-area flakes adopt the golden-yellow hue of bulk gold as described by the Drude-Lorentz-model [128].

Owing to their unique properties, high-aspect-ratio gold flakes have been widely utilized across various fields. In the following we will highlight a few of these sorted for topics and backed up with more comprehensive tables: while most of applications lie in photonics and related fields **(Table 1)** they have also been utilized in nanoelectronics **(Table 2),** medical diagnostics and sensing **(Table 3)**. Additionally, these large flakes serve as building blocks for complex nanodevices **(Table 4)**, catalysis **(Table 5)**, and finally other emerging applications **(Table 6)**.

## Photonics and Plasmonics

Gold flakes exhibit close to perfect crystallinity maximizing conductivity and reducing heat generation due to ohmic losses to the theoretical minimum [8, 35, 36]. Their large contact area with any surface is in general an advantage, as homogeneously grown "normal" single gold crystals would be roughly spherical and, therefore, have only a small and difficult to predict contact area with a substrate. Finally, macroscopic crystals are too extended perpendicular to the substrate to be conveniently further structured by standard top-down techniques (in one exception, one has been used for creating plasmonic ridges on top of the crystal [129]). Consequently, gold flakes have been used in a plethora of optics and photonics related research, as summarized in **Table 1**.

Pristine, unstructured gold flakes provide a chemically stable, well-defined environment for investigating fundamental properties of surface plasmon polaritons (SPPs) in 1D (at the edges) and 2D (on the surface). SPPs form when electromagnetic waves with frequencies smaller than, but near a metal's plasma frequency ($\omega_p(Gold) = 13{,}8 \cdot 10^{15}$ Hz) interact with the metals' conduction band electrons, yielding a coupled state between the quasi-free electrons and light [130]. The currents driven with these high frequencies are prone to ohmic losses, in contrast to metals being perfect electrical conductors for low frequencies. To effectively apply plasmons (a word often used instead of SPP), any additional losses due to crystal faults should be kept minimal, an intrinsic feature of monocrystalline gold flakes. Plasmon propagation phenomena have been researched extensively [65, 131, 132] and is a possible method to measure the dielectric function of monocrystalline gold. As gold flakes can be transferred, e.g., via polymer droplet methods [115], also the quantum phenomenon of non-locality has been examined by stacking flakes with extremely thin spacers made from 2D materials [133].

For more sophisticated applications, gold flakes offer an exceptionally well-suited starting substrate for top-down fabrication of well-defined complex geometries via lithographic and/or milling techniques [7, 35, 119-121, 123]. Desired functional elements contain optical antennas, plasmonic waveguide for nanocircuitry systems, and more advanced structures integrating multiple optical and optoelectronic components [25, 36, 119, 122, 123, 134-138].

Structured monocrystalline large-area gold flakes were first reported as an optical material by Wiley *et al.* [139], where a mechanical skiving method was used to cut a flake into quasi 1D plasmonic Fabry-Perot resonators, where plasmons form modes comparable to a guitar string [140]. However, the first use of gold flakes as a basis for fabricating geometrically versatile plasmonic resonators and waveguides was reported by Huang *et al.* [8] (**Figure 5**). Here, e.g., double wire and bow-tie shaped optical antennas[141] were fabricated via Ga FIB milling, realizing antenna gaps of only a few tens of nanometers where light is concentrated to extremely small volumes and several orders of magnitude higher intensities than possible with classical optics, verified via non-linear two photon photoluminescence (TPPL). The superior optical properties of gold flakes were demonstrated in direct comparison to identical geometries made from sputtered gold (**Figure 5**(A<sub>iii</sub>))

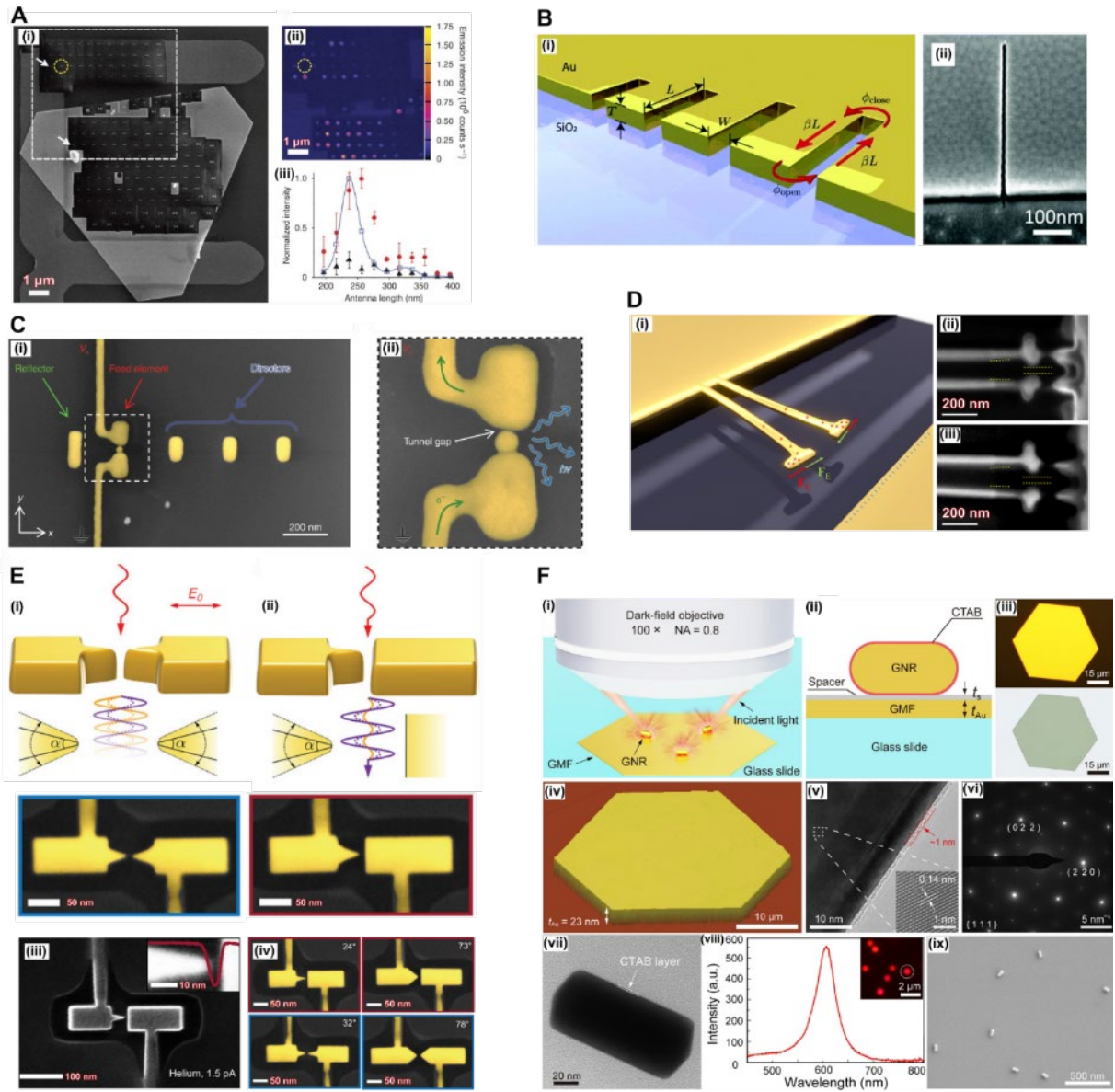

Figure 5: Applications of optical antennas made from monocrystalline gold flakes. (A) Bowtie and linear antennas fabricated in both monocrystalline and polycrystalline gold. (A$_i$) SEM image of linear antennas fabricated from monocrystalline gold flake (middle) with patches of a vapor-deposited polycrystalline gold film (upper portion), with the respective two-photon photo luminescence (TPPL) mapping (in A$_{ii}$). In (A$_{iii}$), the averaged integrated TPPL intensity from linear nanoantennas on monocrystalline (red dots) and polycrystalline (black triangles) gold. Panel A reprinted with permission from Huang *et al.* Ref.[8]. Copyright © 2010, Springer Nature Limited. Panel B shows plasmonic nano slit antennas fabricated by helium ion milling. (B$_i$) Schematic visualization with corresponding SEM at (B$_{ii}$). Panel B Reproduced from Chen *et al.* [142] with permission from the Royal Society of Chemistry. (C) Plasmonic Yagi-Uda antenna driven by inelastic electron tunneling. (C$_i$) SEM image of the antenna containing a reflector, a feed element with kinked connectors, and three directors fabricated via Ga-FIB on glass. A magnified view of the feed element in (C$_{ii}$) (dashed white rectangle) reveals an asymmetrically positioned particle attached via dielectrophoresis, forming a tunnel gap toward the top antenna arm. Panel C adapted from Ref. [119]. Copyright © 2020, The Author(s). (D) An electromechanically tunable suspended nanoantenna. A 3D schematic of the structure is depicted in (D$_i$), while SEM images in (D$_{ii}$) and (D$_{iii}$) illustrate how the gap width expands from 40 nm to 70 nm as the applied voltage increases from 0 V to 20 V. Panel D Reprinted (adapted) with permission from Chen *et al.* [123]. Copyright © 2016, American Chemical Society. (E) Effect of local symmetry breaking on the second harmonic generation (SHG) process in gold nanoantennas. Upon excitation with a linearly polarized laser at frequency $\omega$, the SH efficiency depends on the gap geometry. (E$_i$)-(E$_{ii}$) Symmetric-gap antenna and asymmetric-gap antenna respectively with colored SEM images. Both antennas have a gap size of 9 nm. (E$_{iii}$) shows

an example of the minimum achievable gap size of < 5 nm. The red line displays the line profile along the center of the SEM image. (E$_{iv}$) tuning SHG by varying the degree of local symmetry breaking. Panel E reprinted from Meier *et al.* [143]. Copyright: © 2023 The Authors. Advanced Optical Materials published by Wiley-VCH GmbH. (F) highlights the integration of gold nanorods (GNRs) with gold micro flakes (GMFs). (F$_i$) and (F$_{ii}$) provide a schematic visualization of the design and its cross-section, while (F$_{iii}$) presents optical microscopy images in both reflected and transmitted modes. The structural characteristics of the gold flake are further detailed through atomic force microscopy in (F$_{iv}$) and TEM imaging in (F$_v$), which includes a high-resolution zoom-in and the corresponding electron diffraction pattern shown in (F$_{vi}$). In (F$_{vii}$), a TEM image of a GNR. (F$_{vii}$) Scattering spectrum of a GNR on a glass. Inset: dark-field scattering image of the measured GNR (circled). (F$_{ix}$) shows a SEM image of assembled system of GNR on gold flake. Panel F Reprinted (adapted) with permission from Liu *et al.* [144]. Copyright © 2022, American Chemical Society.

Optical accessible plasmonic resonances can also form in slits carved into a gold flakes edge[142]. With single-digit nanometer widths fabricated by means of focused Helium ion beam milling structures with multiple overlapping resonances are realized (**Figure 5**(B)), allowing for quantum properties of localized plasmons to be assessed.

The structural properties of gold flakes allow the realization of long smooth wires using ion-beam milling [8]. This allows for electrically connecting plasmonic optical antennas with undisturbed plasmonic resonances [145]. Based on this Kern *et al.* [146] used dielectrophoresis to trap a nano particle in the antenna gap, allowing the excitation of the plasmon resonance via inelastic electron tunneling, finally leading to spectrally shaped light emission. As a follow-up development, Kullock *et al.* [119] to demonstrate highly directive, compact, and electrically driven Yagi-Uda antennas (**Figure 5**(C)), offering a potential low-footprint solution for chip-based photonic communication applications. Another idea has been shown by Chen *et al.* [123] suspending a connected optical nanoantenna in air so that the gap between the antennas arms can be expanded by charging the antenna arms, as shown in **Figure 5**(D). This electromechanically tunable device has possible applications in optical nanoelectromechanical systems (NEMS).

Pushing light localization to the extreme, Helium ion microscopy (HIM) allows to realize asymmetric gaps smaller than 5 nm with tip radii as low as 8 nm (**Figure 5**(E)) to optimize non-linear optical effects like second harmonic generation (SHG). Smaller gaps can be realized only by (nano) particle on a mirror ((n)POM) designs [147] where gold flakes can serve as the optimal mirror [144, 148]. This configuration facilitates extreme optical confinement in a thin layer between the flake and nanospheres or rods (**Figure 5**(F))[144]. Similarly, silver nanoparticles on gold flakes have been used [148].

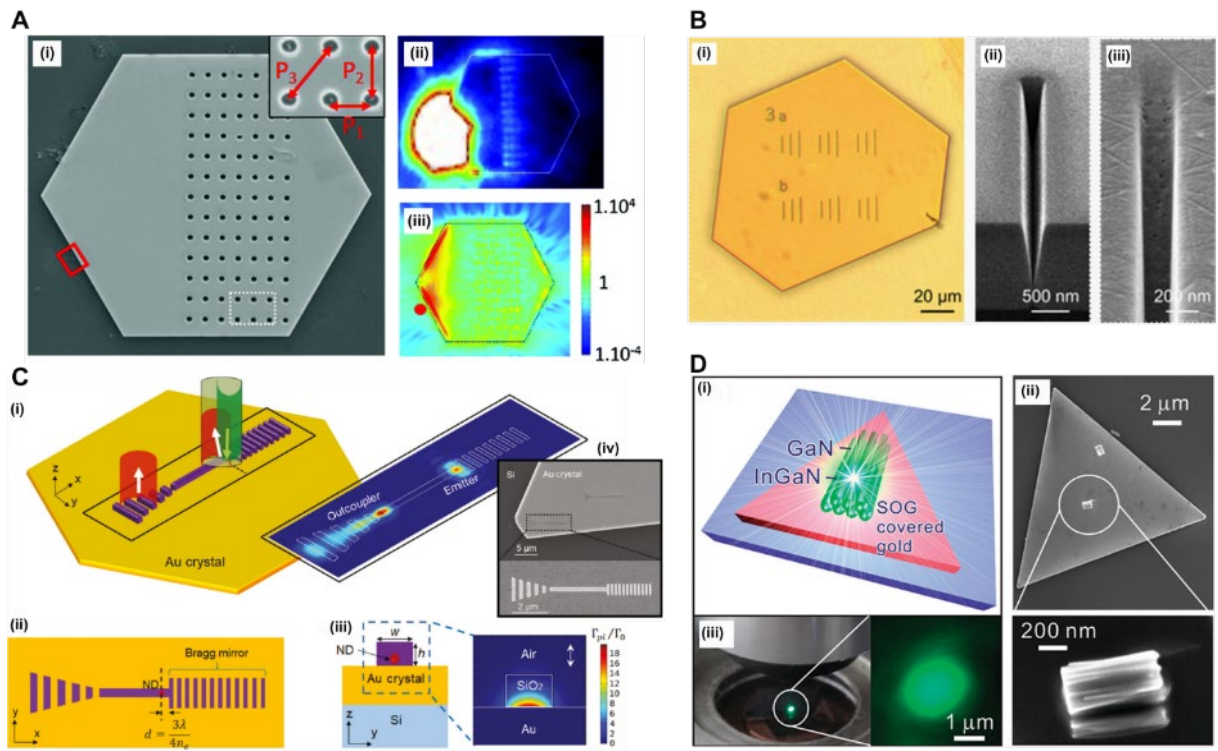

Figure 6: Monocrystalline gold flakes for light-matter interaction. (A) Plasmonic hole array fabricated *via* FIB milling on a monocrystalline gold flake to investigate plasmon propagation excited by a single photon emitter. (A$_i$) SEM image of the milled structure, with the red rectangle indicating the quantum nano emitter's location and highlighting key geometrical parameters corresponding to the white dashed rectangle. (A$_{ii}$) Experimental and (A$_{iii}$) simulated surface plasmon signal transfer maps on the gold flake surface. Panel A reproduced from Kumar *et al.* [149] with permission from the Royal Society of Chemistry (B) Plasmon polariton channel's structure to couple to single fluorescent molecules. (B$_i$) shows a microscope image of a gold flake containing V-grooves of varying sizes. (B$_{ii}$) Helium ion microscopy image showing a cross-section of a V-groove alongside a nano mirror. (B$_{iii}$) Zoomed-in view of the nano mirror at one end of the V-groove. Panel B reprinted (adapted) with permission from Kumar *et al.* [150]. Copyright © 2020 American Chemical Society. (C) Single-photon emitter platform based on a structured dielectric on top of a flake. (C$_i$) and (C$_{ii}$) represent a schematic of the device layout and working principle. Dielectric nano ridges were fabricated by EBL atop a gold flake. The simulated far-field image for the coupled system is shown on the right of (C$_i$). (C$_{iii}$) Cross-section visualization of the system, including mode profile, indicating the distribution of Purcell enhancement. (C$_{iv}$) SEM image of the fabricated device on gold crystal. Panel C adapted from Siampour *et al.*[151]. (D) Gold flake-based green nano laser (spaser). (D$_i$) Schematic illustration of the lasing architecture, comprising a bundle of green-emitting semiconductor nanorods coupled to an underlying monocrystalline gold flake. (D$_{ii}$) FE-SEM image of the hybrid system, with a magnified view detailing the InGaN/GaN nanorod bundle positioned atop the gold flake. (D$_{iii}$) Emission of green laser light from the hybrid structure. Reprinted (adapted) from Wu *et al.* [135]. Copyright © 2011 American Chemical Society.

Examples of single emitters coupled to (structured) gold flakes are displayed in **Figure 6.** Also termed hybrid optical antenna structures, they have been realized using fluorescent molecules, quantum dots or defects in 2D/3D materials. The possibility to achieve a large field enhancement near plasmonic resonators is identical to providing an increased local density of states (LDOS) at the respective position, which enhances the emission rate of excited electronic states [152]. Also known as "Purcell enhancement", there is a classical derivation which relates emission enhancement to large quality factors and small mode volumes of a resonator [153]. Therefore, structures made from monocrystalline gold surpass polycrystalline materials, as both the q-factor and the quantum efficiency, the ratio of originally emitted photons reaching the far field, increase proportionally to the conductivity of the resonator material. For example, the photoluminescence (PL) of single quantum dots coupled to a gold flake in dependence on the thickness of a PMMA spacer layer shows a significant change in the optical properties [124].

Stronger coupling can be achieved by placing the single emitter at the edge of a flake as realized in [149] (**Figure 6(A)**) using a nano diamond with a single nitrogen-vacancy center. In addition, a grating was engraved into the flake acting as a plasmon wavelength filter. An example of a molecular fluorescent photon source uses single dibenzoterrylene (DBT) molecules in anthracene nanocrystals. They have been integrated within grooves carved by FIB into gold flakes (**Figure 6**(B)) [150] showing a 50% emission enhancement and 14 μm plasmon decay length. More sophisticated geometries employ antennas and waveguides towards realizing a full nanophotonic device. A nanostructured dielectric material on top of a gold flake can be designed to realize waveguiding modes, Bragg mirrors and optical antennas [151] (**Figure 6**(C)). This allows for structuring around an embedded single emitter, e.g., a nano diamond containing an GeV-(Germanium-vacancy)-center as a single photon source that is enhanced 15-fold.

A potentially disruptive application of plasmon hybrid devices is spasing [154] which is in resembling the laser the abbreviation for Surface Plasmon Amplification by Stimulated Emission of Radiation. Sub-diffraction-limited laser operation in the green spectral region has been demonstrated in a hybrid metal-oxide-semiconductor (MOS) plasmonic nanocavity structure [135] (**Figure 6** (D)), again benefitting greatly by the low losses within gold flakes lowering the laser threshold. Additionally, monocrystalline gold-based metasurfaces enabled precise control over the anisotropic and isotropic contributions to second harmonic generation (SHG) near localized surface plasmon resonance conditions [155, 156], making them valuable for studying nonlinear optical phenomena in metal thin films [77, 157], as well as nanophotonic imaging and probing of low-dimensional materials [158].

Beyond FIB milling, various other approaches to fabricate monocrystalline nanostructures benefit from large gold flakes, including electron beam lithography (EBL), direct e-beam writing (DEBW), and nanoscale skiving with ultramicrotome [121, 122, 139, 151]. To realize an array of optical antennas spanning a whole gold flake Méjard *et al.* [121] employed standard electron-beam lithography (EBL), followed by dry etching (**Figure 7(A)**) improving scalability compared to FIB milling.

Plasmonic circuits make use of the light localization in two dimensions but elongate the third dimension to realize wave guides with deep subwavelength cross-sections. Also, the plasmon wavelength is (possibly much) smaller than the vacuum wavelength of light with a given frequency ($\lambda_p < \lambda$) [140]. Using the large area of a gold flake, the fabrication of a vast amount of single wire plasmonic waveguides with randomly changing but never repeating lengths (**Figure 7(B)**) has advanced the quantitative understanding of plasmon propagation [159]. The reproducibility of the wire cross sections and end cap geometry as well as the always identical crystallinity allowed to reliably model the structures analytically as Fabry–Pérot resonators [160].

Based on former fundamental research [161], Krauss *et al.* [162] utilized gold flakes to fabricate a multimode plasmonic nanocircuit composed of forked two-wire transmission lines that can be used to reversibly mapping and sorting photon spin states at the nanoscale (**Figure 7**(C)). The device relies on a symmetric and antisymmetric eigenmode that interfere in carefully fabricated bent waveguides. This control over the spin angular momentum of light could pave the way for innovations in quantum information processing and spin-based photonic technologies. **Figure 7**(D) shows instead a waveguide made from a chain of slit resonators, with their distances carefully tuned in a way that a topologically protected plasmonic edge state emerges from the interference of all reflections and transmissions that can be measured via photo electron emission microscopy (PEEM) [163].

Furthermore, gold flakes have been widely studied as a model system for exploring fundamental optical properties in monocrystalline thin films, including strong vibrational coupling in ultra-high frequency plasmonic nano resonators [134], observing plasmonic skyrmion dynamics with deep subwavelength resolution [136], revealing quantum-mechanical effects in the luminescence emanating from thin monocrystalline gold flakes [137] and detecting the plasmon-polariton quantum wave packet [164].

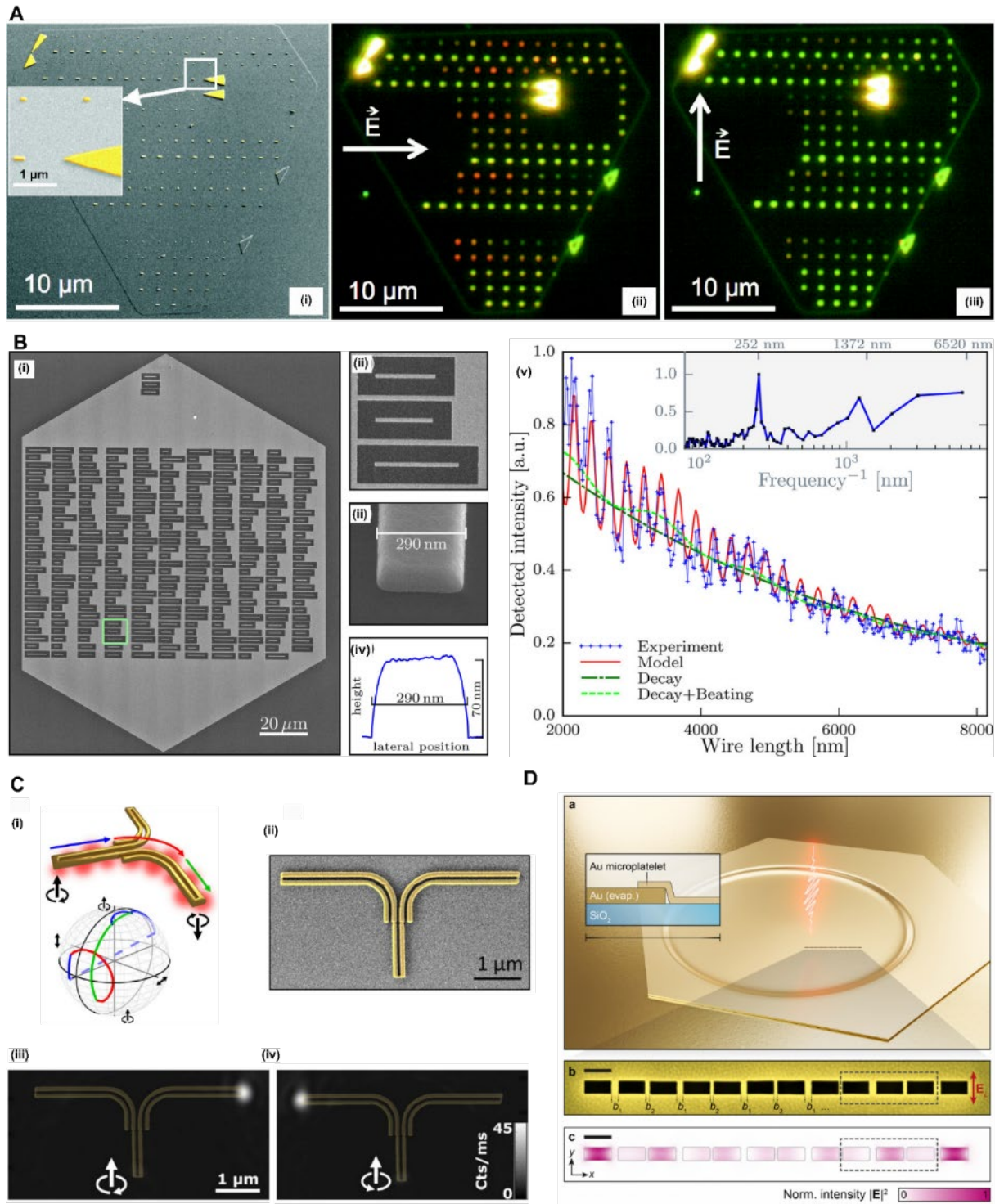

**Figure 7: Non-FIB fabrication and plasmonic circuits.** (A) Shows a fabricated monocrystalline nanoantenna multiple arrays. ($A_i$) presents an SEM image of nanoantenna rods sculpted from a large monocrystalline gold flake, with the inset providing a magnified view. ($A_{ii}$) and ($A_{iii}$) display dark-field images captured under polarization aligned longitudinally and transversely to the rods, respectively. In ($A_{ii}$), the color variations indicate a spectral redshift in the plasmonic resonance as the antenna length increases. Conversely, in ($A_{iii}$), the color remains uniformly distributed, as the transverse polarization interacts with the short axis of the nanorods, which exhibits minimal variation across the array. The bright triangular features serve as alignment landmarks. Panel A adapted from MÉJARD *et al.* Ref. [121]. Copyright © 2017, Optical Society of America. (B) Fabricated plasmonic circuitry arrays for plasmon transmission analytical analysis. ($B_i$) SEM image of the full fabricated arrays within a gold flake, with higher-magnification views shown in ($B_{ii}$) and ($B_{iii}$). Panel ($B_{iv}$) shows AFM profile of the wire cross-section.

(B$_v$) demonstrate the detected signals from the plasmonic circuitry arrays, presented alongside the proposed Fabry–Pérot model. Reprinted (adapted) with permission from Geisler *et al.* [159]. Copyright © 2017 American Chemical Society. (C) Spin-Optical Nanodevice. (C$_i$) Sketch of the spin-optical nanodevice, where plasmons first propagate linearly (blue), then follow a curved path (red), and finally continue with another linear segment (green). The device functions similarly to an electron spin transistor, with in-coupling (source), out-coupling (drain), and a central gate region. In this gate region, the spin state is defined by the accumulated phase, acting like a gate voltage. The lower part shows the trajectory of the photon's pseudospin state throughout the device, represented on a Poincaré sphere. (C$_{ii}$) colored SEM-image. (C$_{iii}$) and (C$_{iv}$) shows a CCD-images for excitation of the left-handed (C$_{iii}$) or right-handed (C$_{iv}$) pseudospin state with the analyzer in the detection path set to transmit only emission of the left-handed or right-handed pseudospin state. The structure's position is indicated by the overlaid colored SEM-image. Reprinted (adapted) with permission from Krauss *et al.*[165]. Copyright © 2019, American Chemical Society. (D) Plasmonic nano slit Su–Schrieffer–Heeger (SSH) chain with nontrivial topology: (D$_a$), Chains of plasmonic nano slit resonators written into a monocrystalline Au micro flake using helium focused ion-beam milling. The micro flake covers a hole in a vapor-deposited Au film residing directly on a smooth glass substrate. (D$_b$), Top-view scanning electron microscopy (SEM) image (false color) of a nontrivial nanoslit SSH chain consisting of twelve coupled resonators separated by bridges with alternating widths, $b_1$ and $b_2$, as indicated. (D$_c$), Simulated near-field intensity of a mid-gap mode (COMSOL) exhibiting localized nearfield intensity at the two outermost nanoslits. Reprinted from Schurr *et al.* [163]. Copyright © 2025, the authors.

**Table 1:** Gold flake applications in nanophononics

| Method | Application summary | Ref. |
|---|---|---|
| As synthesized | Fundamental properties | [25, 77, 136, 137, 156, 157, 166-170]. |
| | Surface plasmons | [34, 132, 171-174]. |
| | Edge plasmons | [65]. |
| | Quantum emitter enhancement | [124, 149]. |
| | Strong vibrational coupling | [134]. |
| | As substrate/mirror | [144, 148, 175]. |
| | Thinning | [36]. |
| | LSPR properties and control | [25, 170, 176, 177]. |
| Structured: resonators | Rods | [8, 139, 178-188]. |
| | Slits | [142, 164, 189-191]. |
| | Evolutionary optimized | [192]. |
| | Electrically connected rods | [119, 123, 145, 146, 193-199]. |
| Structured: waveguides | Fundamental properties | [159, 161, 162, 200-202]. |
| | Non-linear optics | [169, 203, 204]. |
| | Quantum emitter enhancement | [150, 191, 205]. |
| Structured: full devices | Receiver & emitter | [206]. |
| | Quantum emitter enhancement | [151]. |
| Structured: gratings/meta surfaces | Fundamentals | [53, 120, 138, 207]. |
| | Focusing | [131]. |
| | Sensing | [208]. |
| | Non-linear | [155, 209, 210]. |
| Flake hybrids | Phonon-polaritons | [135, 158, 211]. |
| | Non-locality | [133]. |

Note: "As synthesized" means gold flake without further fabrication or modification

## Electronics

High-aspect-ratio gold flakes have been utilized also in various nanoelectronics applications, showing, e.g., potential for high-dielectric-constant materials, printable electronics, capacitors, and electrodes. Li *et al.* [212] introduced a hybrid nanocomposite of gold flakes and fibrils with tunable conductivity, ranging from insulating-like behavior to values approaching pure gold conductivity that can be employed as a humidity sensor (**Figure 8(A)**). In a similar approach gold flake–chitin nanofiber hybrids have been developed by Chen *et al.* [213], with possible applications in humidity sensing, breath analysis, and pressure sensing evaluation via electric conductivity of the hybrid circuit. Such a system could be used for speech recognition, health monitoring, or respiratory analysis, depending on its design and sensitivity.

Gold flakes have also been explored as contact and electrode materials in electronic circuits [10, 214-217]. Moon *et al.* [10] proposed multilayered gold flakes as a novel stretchable electrode material suitable for organic-based electronic devices (**Figure 8(B)**). These electrodes exhibited excellent electrical stability under repeated stretching cycles, highlighting their potential for printable and wearable sensing applications [28, 218]. Zhu *et al.* [217]. reported a plasmonic platform based on nanowires fabricated from synthesized gold flakes that have been treated by electromigration to form irregular nanometer sized gaps for photovoltaic and electroluminescence applications. Seo *et al.* [215] demonstrated a flexible resistive switching memory device using an ultrathin composite film of gold flake nanosheets as electrodes (**Figure 8(C)**), highlighting its potential for wearable applications due to its paper-like mechanical flexibility. Boya *et al.* [214] integrated synthesized gold flakes as top contact electrodes in large-area metal–molecule–metal electrode systems for the electrical analysis of molecules. Similarly, gold flakes were used to establish Ohmic top contacts for vertically grown nanowires of uneven height, leveraging electromigration of the (111) surface towards the shorter nanowires, as shown in **Figure 8(D)** [216].

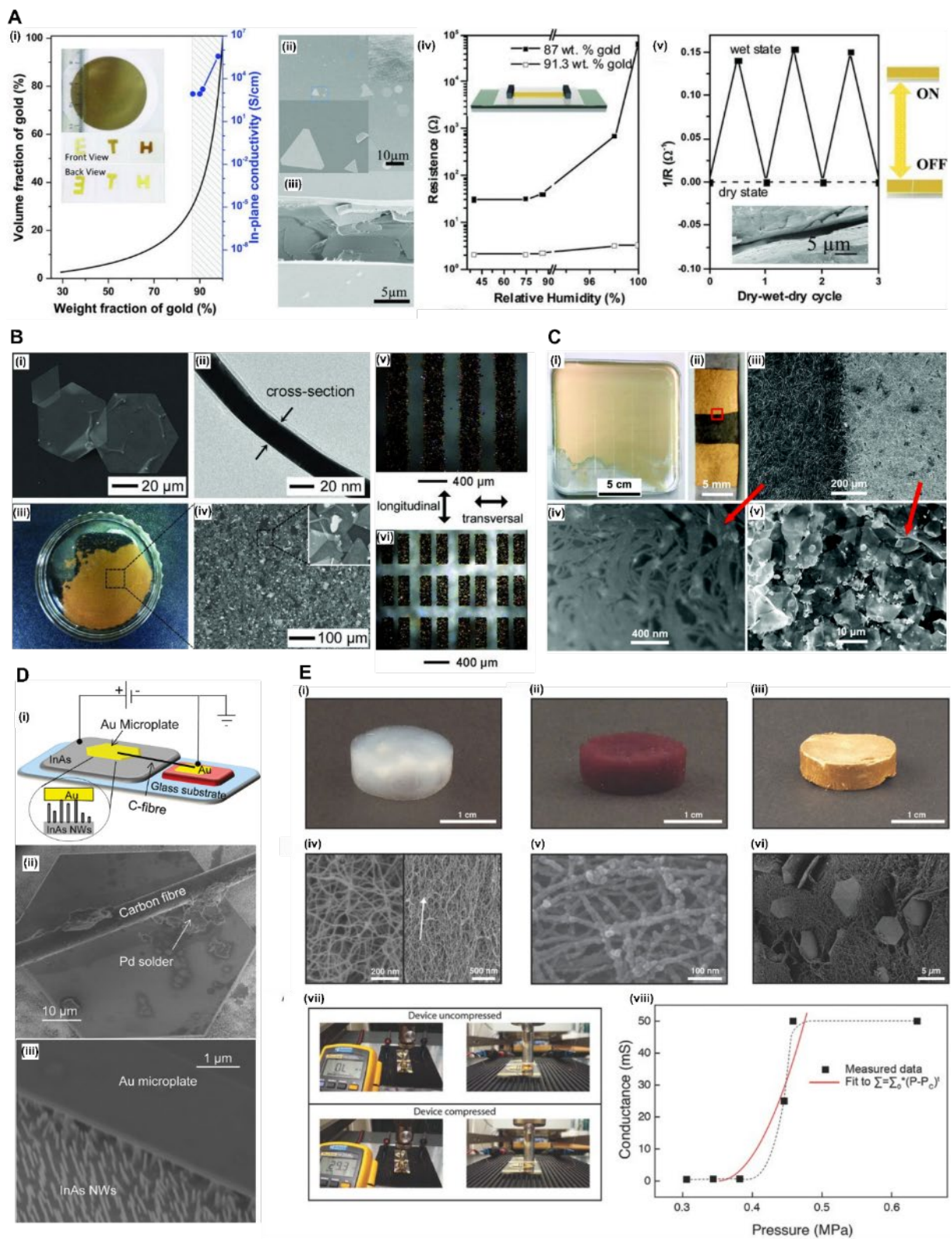

**Figure 8: Monocrystalline gold flakes as electronic building blocks.** (A) Applications of Gold flake amyloid fibril hybrid. (A$_i$) Correlation between gold content in hybrid films and their conductivity. The inset shows a photograph of typical hybrid films. SEM images of the hybrid films depicting (A$_{ii}$) surface morphology and (A$_{iii}$) fracture section. (A$_{iv}$) and (A$_v$) Resistance-based humidity sensor response for films with varying gold flake compositions. The first inset illustrates the device, the second a typical micro-scratch. Panel A reprinted with permission from Li *et al.* [212]. Copyright © 2013 WILEY-VCH Verlag GmbH & Co. KGaA, Weinheim. (B) Stretchable patterned gold flake-based electrode. (B$_i$) SEM image of the synthesized gold flakes. (B$_{ii}$) Cross-section TEM image of a gold flake. (B$_{iii}$)

Gold flake film assembled on a water surface. (B_iv) SEM image of the assembled Au flake film; the inset provides a magnified view showing overlapping Au flakes. (B_v)-(B_vi) Optical microscopy images of patterned gold flake electrodes. Panel B reprinted with permission from Moon *et al.* [10]. Copyright © 2013 WILEY-VCH Verlag GmbH & Co. KGaA, Weinheim. (C) Preparation of a flexible gold flake-based device. (C_i) An assembled gold flake film on water. (C_ii) Au flakes electrodes. (C_iii) SEM image of the red square in (C_ii). (C_iv)-(C_v) The enlarged SEM images of electrode structures emphasizing the gold flakes in (C_v) panel. Panel C Reproduced from Seo *et al.* [215] with permission from the Royal Society of Chemistry. (D) Gold micro flake device. Panel (D_i) shows the configuration of Au microplate position. (D_ii) SEM image of the actual assembled device. The Pd metal between the carbon fiber and the microplate is also visible. (D_iii) Zoom in to the nanowire structures beneath the gold flake along the edges. Panel D reprinted (adapted) with permission from Radha *et al.* [216]. Copyright © 2012 American Chemical Society. (E) Gold flake hybrid aerogels. (E_i)-(E_iii) Photographs of (E_i) a standard aerogel, (E_ii) a gold nanoparticle-amyloid aerogel, and (E_iii) a flake-amyloid aerogel. (E_iv)-(E_vi) SEM images of (E_iv) the amyloid aerogel, (E_v) the gold nanoparticle-amyloid aerogel, and (E_vi) the gold flake-amyloid aerogel. (E_vii) Photographs of a pressure sensor device incorporating gold flake-based aerogels in its uncompressed (top) and compressed (bottom) states. (E_viii) Conductance of the gold flake-amyloid aerogel as a function of applied pressure. Panel E reprinted with permission from Nyström *et al.* [219]. Copyright © 2016 WILEY-VCH Verlag GmbH & Co. KGaA, Weinheim.

In another example, biosynthesized gold triangles were assembled onto various substrates as building blocks in thin films for organic vapor sensing, where their resistance decreased upon exposure to weakly polar molecules [220]. This behavior suggests that an analyte containing molecular dipoles influences conductivity, enabling possible chemical sensing applications [221]. Similarly, Ankamwar *et al.* [108] found that an analytes polarity plays a crucial role in determining film resistance. Nyström *et al.* [219] reported the development of ultralow-density amyloid fibril-based aerogels functionalized with gold flakes, which were employed as pressure sensors (**Figure 8**(E))as their conductance varied as a function of the applied pressure. Finally, Zhang *et al.* [12] demonstrated the integration of gold flakes into electrical circuitry by functionalizing an Au triangle–chitosan matrix with an immobilized enzyme capable of glucose detection, monitored through changes in its cyclic voltammogram. A summary of these and more gold flake applications in electronics is shown in **Table** 2.

**Table 2: Au flakes as electronic building blocks**

| Method | Application summary | Ref. |
|---|---|---|
| As synthesized | Building block in electrocatalytic electrodes | [82, 222-225]. |
| | Electrical circuit building blocks | [10]. |
| | Flexible device for resistive switching | [215]. |
| | Nanocomposite with tunable conduction | [212]. |
| | Electrically conductive coatings for vapor sensing | [108, 220]. |
| | Gold flake-based aerogel to conductance pressure sensor | [219]. |
| | Chitin nanofiber hybrid film sensor for humidity, pressure, and more. | [213, 225]. |
| | Strain control electrical devices | [42]. |
| Functionalized flake | Electrical circuit building blocks | [12, 214, 216, 226]. |
| Structured | Open-circuit photovoltaic device | [217]. |

## Sensing

The versatility of gold flakes facilitates their seamless integration into sensing and biomedical applications [12, 35, 53, 107, 108, 212, 213, 219, 220, 225, 227-238]. For example, surface-enhanced Raman scattering (SERS) is a powerful spectroscopic technique that amplifies Raman signals through localized surface plasmon resonances (LSPRs) in noble metals – most notably gold [19, 239, 240]. Raman spectroscopy detects vibrational modes of solids and molecules via inelastic scattering of light, but with a factor of $10^6$ less efficiency than elastic scattering. However, since Raman signals scale non-linearly with the local electromagnetic field, well designed plasmonic resonances enable highly sensitive molecular detection [239, 240]. Gold flakes were employed by Sweedan *et al.*[35] to fabricate an evolutionary-optimized plasmonic sensor for SERS applications (**Figure 9**(A)) working with a wide range of analytes across different states of matter, including nucleotides and proteins [35]. Lv *et al.* [53] reported obtaining huge gold flake synthesized biologically via silk fibroin ((**Figure 9**(B)) as a platform to grow nanotips, generating random hotspots suitable for SERS applications, e.g., to monitor catalytic reactions Lin *et al.*[238] harnessed a doppler grating platform to sense environmental index changes (**Figure 9**(C)). In an alternative approach reported by Xia *et al.* [227] silver nanocubes (AgNCs) were deposited onto a flake surface to fabricate a particle on mirror sensors for SERS (**Figure 9**(D)). Chen *et al.* [233] developed a large-scale, two-dimensional, flexible SERS-based functional platform by growing gold nanoparticles (AuNPs) in situ on the flake surface. Furthermore, Wang *et al.* [235] advanced mono-metal epitaxial growth by fabricating AuNP/Au flake hybrid structures featuring ultrasound-induced hollow gaps within the flakes, significantly enhancing their potential for surface-enhanced Raman scattering (SERS) applications (**Figure 9**(E)).

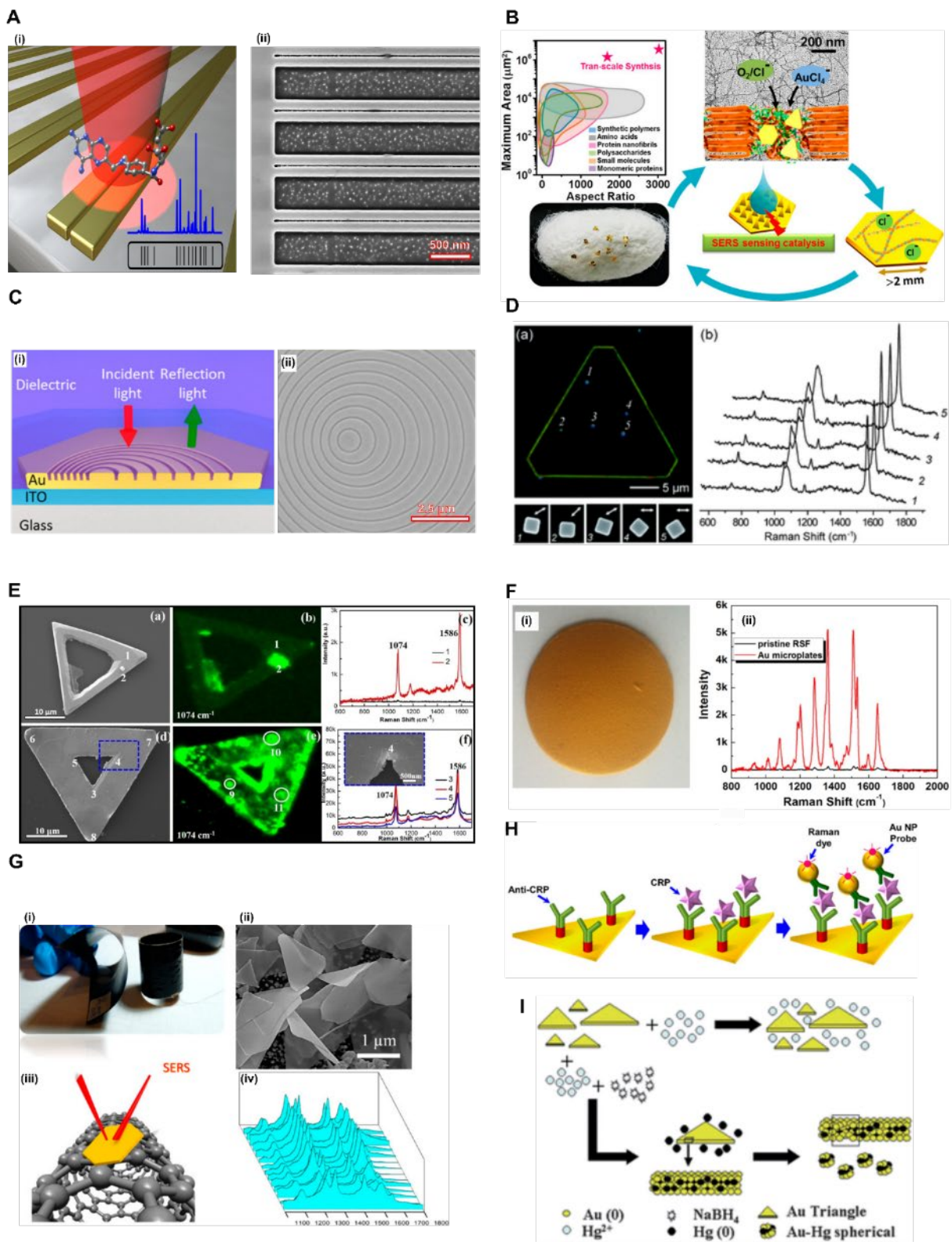

**Figure 9: Monocrystalline gold flakes sensing applications.** (A) Double resonator sensing platform: (A$_i$)Geometry of the SERS sensing platforms. (A$_{ii}$) SEM images of the FIB-fabricated monocrystalline gold double gratings. Reprinted from Sweedan *et al.* [35]. Copyright © 2024 The Authors. Small published by Wiley-VCH GmbH. (B) Plasmonic Doppler grating (PDG) index sensor. (B) Au flakes with lateral dimensions exceeding 2 mm for SERS

applications, synthesized using natural fibrous proteins. Reprinted (adapted) with permission from Lv *et al.* [53]. Copyright © 2018 American Chemical Society. ($C_i$) A schematic showing the structure of a PDG sensor and the configuration of optical setup. ($C_{ii}$) SEM image of a PDG fabricated on the surface of a monocrystalline gold flake for broad range index sensing. Reprinted (adapted) with permission from Lin *et al.* [238]. Copyright © 2019 American Chemical Society. (D) Gold flake-silver nano cube sensing platform: ($D_a$) Optical micrograph of five different silver nano cube (AgNCs) on Au flake. ($H_b$) SERS spectra recorded from the five AgNCs shown in ($D_a$). Panel D Reproduced from Xia *et al.*[227] with permission from the Royal Society of Chemistry. (E) Hollow gold flake sensing platform: ($E_a$), ($E_d$) SEM images of the platform. ($E_b$), ($E_e$) Raman mapping of the analyte signal. Reprinted from [235]. Copyright © 2020, Springer-Verlag GmbH Germany, part of Springer Nature. (F). Au flake hybrid material sensing: ($F_i$) Photo of Au flake/silk nanofibrils composite. ($F_{ii}$) Raman spectra of the analyte. Panel F reprinted from Fang *et al.* [231]. Copyright © 2016 with permission from Elsevier B.V. All rights reserved. (G). Gold flake-carbon nanotube (CNT) composite sensor. ($G_i$) Image of a CNT sheet. ($G_{ii}$) SEM of Au flake -CNT sheet. ($G_{iii}$) Schematic illustration of Raman signal enhancement. ($G_{iv}$) Reproducibility evaluation of CNT sheet–Au flake as SERS substrate. Reprinted (adapted) with permission from Xin *et al.* [232]. Copyright © 2017 American Chemical Society. (H) C-reactive protein (CRP) analyte gold flake immunosensor schematically illustrated. Reprinted (adapted) with permission from Hwang *et al.* [234]. Copyright © 2019, American Chemical Society. (I) Gold flake-based sensor for detecting femtomolar concentrations of mercuric ions, utilizing plasmonic changes induced by reaction with mercury. Panel I Reproduced from Singh *et al.* [228] with permission from the Royal Society of Chemistry.

Hotspots for SERS applications were also created by leveraging the sharp tips and edges of triangular gold flakes, either on rigid flat substrates formed from dried drop-cast films [229, 230], or through introducing Au flakes into hybrid nanocomposites. These included natural 3D matrices such as silk nanofibrils [231], chitin [213], and amyloid fibrils[219], which served as scaffolds for synthesis and immobilization, generating randomly scattered gold flakes within the matrices film, as shown in (**Figure 9**(F)) [231]. Similarly, synthetic 3D matrices, such as carbon nanotubes (CNTs) (**Figure 9**(G)), were employed to fabricate a flexible, gold flake-based SERS substrate [232], exhibiting superior SERS performance compared to its AuNP–CNT counterpart.

Functionalizing large-area gold flakes with molecular recognition elements [241] enables the selective detection of specific analytes from complex mixtures, facilitating the isolation of the desired signal from unwanted background noise [234, 236]. Hwang *et al.* [234] reported a functional SERS-biosensor for the trace detection of protein biomarkers. The platform was achieved by functionalizing the gold flake surface with specific recognition antibodies, thereby enhancing the sensor's specificity, as illustrated schematically in **Figure 9**(H). Similarly, gold flakes were successfully used to detect cancer cells in serum samples [237], macromolecules, including ovarian cancer biomarkers [236], as well as monoatomic ions, such as mercury, where optical monitoring after chemical reduction of the ions allowed detection of their concentration, as demonstrated by Singh *et al.* [228] (**Figure 9**(I)).

**Table 3: Integrating Au flakes into sensing platforms.**

| Method | Application summary | Ref. |
|---|---|---|
| As synthesized | SERS substrate obtained by drop-casting and drying of Au flakes | [229, 230] |
| | Au flakes synthesized and embedded within nanofibril matrix | [231] |
| | Au flakes within CNT sheet flexible matrix | [232] |
| | Possible NIR-absorbing or antennas for hyperthermia of cancer cells | [107] |
| | Optical plasmonic change of Au flake upon reaction with mercury | [228]. |
| Functionalized flakes | Functionalized with antibodies for selective SERS sensing | [234, 236]. |
| | Folic acid conjugated for cancer detection | [237]. |
| | Synthesizing nanotip structures on the flake surface for SERS | [53]. |
| | Gold nanoparticles were grown on flake surface for SERS | [233]. |
| | AgNCs deposition on the surface of Au flake for SERS sensing | [227]. |
| Structured | Doppler grating based environmental index sensors | [238]. |
| | Evolutionary optimized SERS sensing platform | [35]. |
| | AuNPs grown on the flake with ultrasound-induced hollow gaps | [235]. |

Note: "As synthesized" means gold flake without further fabrication or modification

## Scanning probe microscopy

The atomically flat surface of large-area gold flakes is optically accessible from the front and from the back side due to their partial transparency originating from their nanometric thickness [8, 36, 242]. This constitutes an ideal substrate for various scanning probe microscopy (SPM) techniques, like scanning tunneling microscopy (STM) [242, 243] and tip-enhanced Raman spectroscopy (TERS) [244-246]. They allow for investigations of adsorbed species down to the single molecule resolution, providing insights into binding sites and molecular orientations, and can additionally benefit from the chemical information using tip-enhanced Raman spectroscopy (TERS; seminal paper, not using flakes: [247]).

Gold flakes used as TERS or STM substrates benefit greatly from the gap-mode emerging between the tip and the gold surface obtained by bringing the tip physically in close proximity to the gold surface as shown schematically in **Figure 10**(A-B). Ren *et al.* [244] studied thiol molecules adsorbed in self-assembled (sub)monolayers, while Pettinger *et al.* [248] (**Figure 10**(A)) used TERS to investigate several different analyte species physisorbed on gold flakes.. Deckert *et al.* [245] utilized flat crystalline gold flakes for label-free investigations of biomolecules using a back-reflection geometry setup, as depicted in **Figure 10(C)**). This approach allowed the simultaneous collection of molecular and topological information. In a related study, Pashaee *et al.* [249] employed a back-scattering TERS geometry with a polarized light source to investigate thiolated molecular species, revealing the polarization dependency of the excitation light at the tip/substrate interface. Additionally, Pashaee *et al.* [250] characterized the edges of graphene-like and graphitic platelets composed of a few layers of graphene deposited on a gold flakes substrate (**Figure 10(D)**). The setup also enabled a quantitative investigation of surface plasmon resonance (SPR) and near-field (NF) heating experienced exclusively by molecules directly contributing to the TERS signal [246].

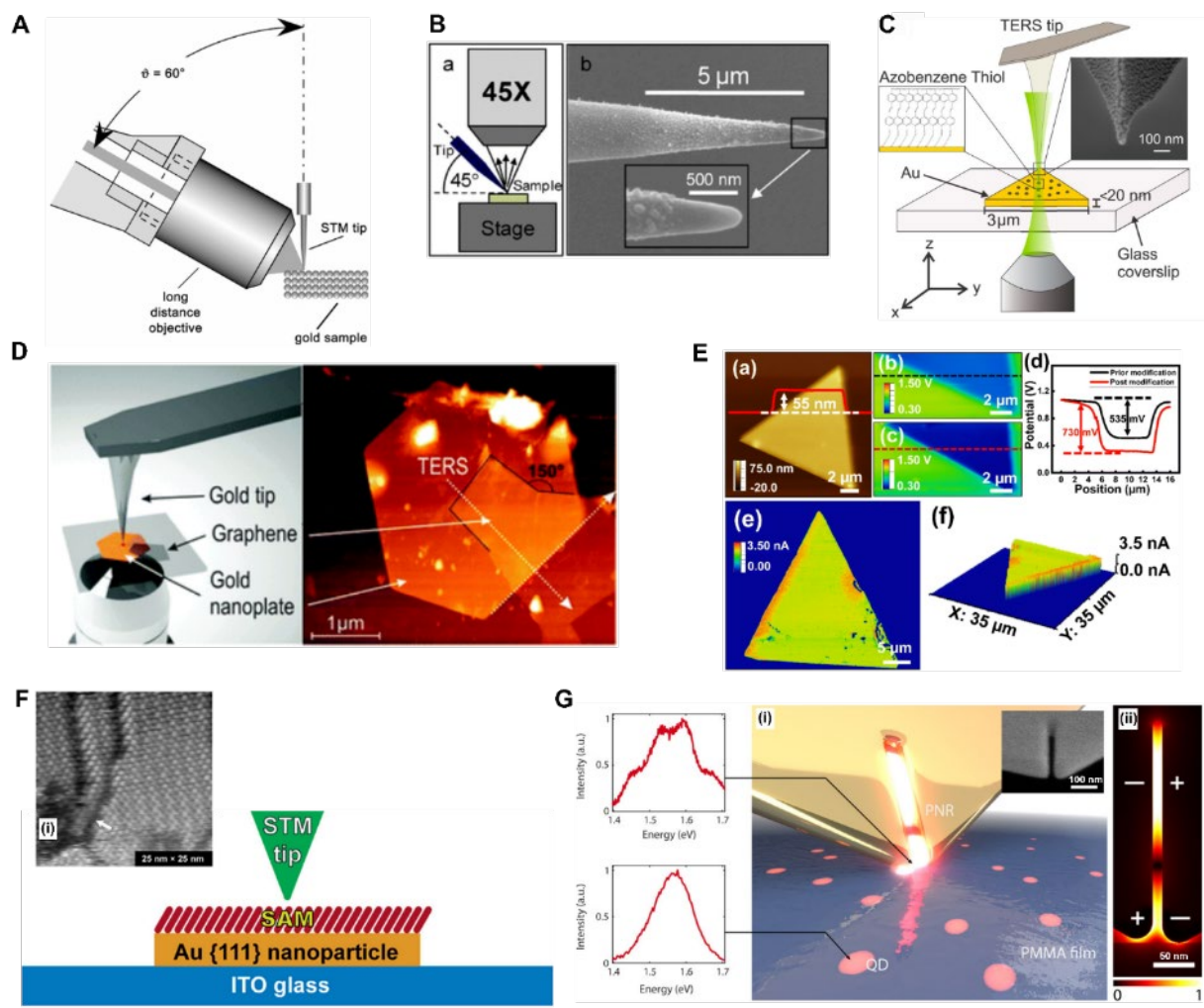

**Figure 10: Au flakes in scanning probe microscopy.** (A)-(C) Different experimental setup arrangements of gold flake platform in TERS and STM. (A) Tip setup using a 60° arrangement. Panel A reprinted with permission from Pettinger et.al [248] © 2004 American Physical Society. ($B_a$) vertical configuration setup. ($B_b$) SEM images of the tip. Panel B reproduced from Pienpinijtham *et al.* [251] with permission from the Royal Society of Chemistry. (C) Setup in back-reflection geometry, showing the gold flake and self-assembled monolayer (SAM). The inset displays an SEM image of the tip. Panel C reprinted (adapted with permission from Pashaee *et al.* [249]. Copyright © 2013, American Chemical Society. (D) TERS-setup illustrating the gold flake and graphene on the left, with AFM images on the right showing a gold nanoplate partially covered with graphene. Panel D reproduced from Pashaee *et.al* [250] with permission from the Royal Society of Chemistry. (E) Electrical characterization studies and tunneling current imaging for Au flake. ($E_a$) AFM topography of triangular Au flake. ($E_b$)- ($E_d$) Surface potential characterization. ($E_e$)- ($E_f$) HD-Kelvin probe force microscopy and contact-AFM characteristics of SAM modified Au flake for electrical output performance. Panel E reproduced from Zhang *et al.* [243] with permission from the Royal Society of Chemistry. (F) Schematic of a Scanning tunneling microscopy (STM) setup and high-resolution imaging of a gold flake surface functionalized with a SAM. The corrugation is the √3 × √3 R30° molecular lattice characteristic of well-ordered alkanethiol SAMs, with dark features (arrow) indicating structural domain boundaries. Panel F reprinted (adapted) with permission from Dahanayaka $et$ $al.$ [242]. Copyright © 2006, American Chemical Society. (G) Strong coupling via precise nano positioning of a gold flake-based resonator probe. ($G_I$) Illustration of the gold flake-based probe interacting with QDs (quantum dots) embedded in a polymer film. Left panel: The spectrum of a QD changes significantly when coupled to the slit-like probe at the tip apex. Inset: SEM image of a nano resonator at the apex of a probe tip. ($G_{II}$) Map of the electric field distribution of the resonator mode used in the experiment. The + and − signs indicate the instantaneous charge distribution highlighting the mode's weakly radiative quadrupolar character. Reprinted (adapted) with permission from Groß $et$ $al.$ [191]. Copyright © 2018, The American Association for the Advancement of Science.

STM studies have highlighted the unique advantages of gold flakes for high-resolution surface analysis. Zhang and colleagues [243] realized an improved friction driven triboelectric generator by utilizing the atomically flat surface together with a self-assembled monolayer (SAM) interacting with an STM tip (**Figure 10**(E)). Similarly, Dahanayaka $et$ $al.$ [242] demonstrated the use of atomically flat gold flakes for SAMs of alkanethiols, capturing structural details at the atomic level by high-resolution STM imaging (Figure 10($F_I$)). It is also possible to realize the TERS tip itself from gold flakes as shown by Groß $et$ $al.$ [191] (**Figure 10**(G)), where FIB-cutting a slit resonator into the gold flake edge after its transfer to an AFM cantilever allowed to establish strong coupling to a mesoscopic colloidal quantum dot. These experiments open the possibility for ultrafast coherent manipulation of the quantum dot–plasmon system under ambient conditions.

## Table 4: Gold flakes for scanning probe microscopy

| Method | Application summary | Ref. |
|---|---|---|
| Functionalized flakes | Flat substrate for STM sample investigation | [242] |
| | Gold flake substrate for TERS-based SAM analysis | [244] |
| | Immobilization substrate for TERS biomolecule investigation | [245] |
| | Study of thiolated molecule species with polarized light | [249] |
| | Quantitative information of the SPR and NF tem. by TERS | [246] |
| | Electrical tunneling current imaging | [243] |
| As synthesized | Substrate for TERS investigating physical absorbates materials | [248, 251] |
| | TERS characterizing graphene-like and graphitic sheets | [250] |
| Structured | Structured gold flake as a probe coupled to a colloidal quantum dot | [191] |

Note: "As synthesized" means gold flake without further fabrication or modification

## Catalysis

Gold nanostructures are well-known for their heterogeneous catalytic activity [219, 252-256], a remarkable property that emerges exclusively at the nanoscale. By precisely controlling the size, shape, and preferential facet orientation of gold nanostructures, their catalytic performance can be significantly enhanced and tailored for specific reactions [254-256]. For instance, Primo *et al.* [255] developed a hybrid catalyst by incorporating large-area gold flakes with a (111) facet orientation into a graphene matrix. This configuration exhibited superior catalytic activity compared to non-oriented gold structures, enabling a range of reactions, including selective oxidations, reductions, and coupling reactions.

In a comparative study, Goyal *et al.* [222] evaluated the catalytic performance of spherical gold nanoparticles and large-area gold flakes in the oxidation of dopamine and ascorbic acid. The gold flakes demonstrated higher catalytic activity, likely due to the large exposed (111) lattice planes. Similarly, gold flakes were assessed in the degradation of azo compounds such as methylene blue, where they showed exceptional catalytic activity (**Figure 11**(A,B)) [257]. Furthermore, He *et al.* [237]  and Li *et al.* [225]  reported that gold flakes effectively catalyze the decomposition of $H_2O_2$ to $O_2$. He *et al.* also reported high efficiency and selectivity in the catalytic hydrogenation of α,β-unsaturated aldehydes, and Momeni *et al.* [223] demonstrated efficient electro-catalysis of formic acid on carbon ionic liquid electrodes modified with gold flakes.

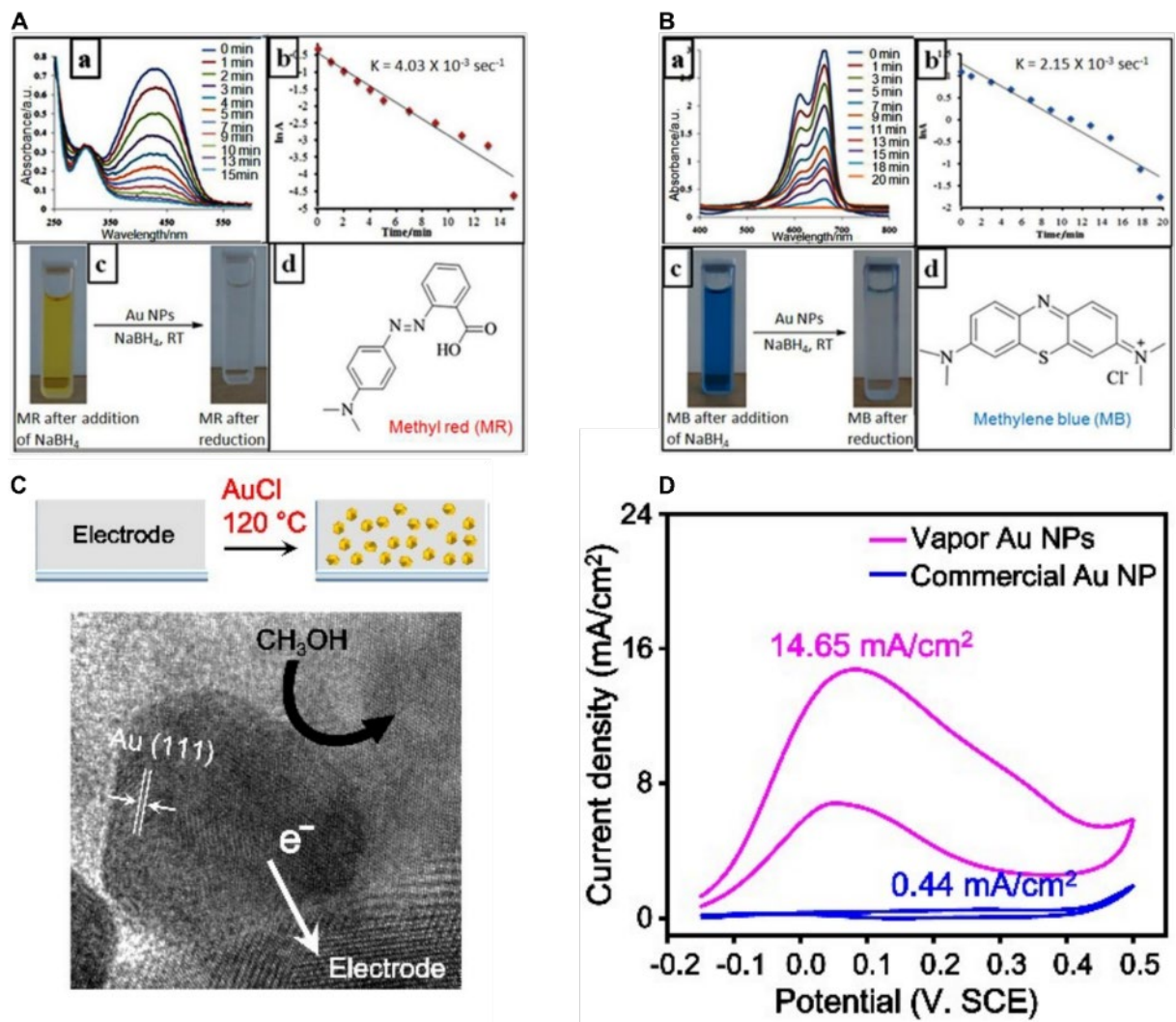

Figure 11: Gold flakes in heterogeneous catalysis. (A)-(B) Catalytic degradation of Methyl Red (MR) in (A) and Methylene Blue (MB) in (B). (A$_a$) and (B$_a$) show UV-visible spectra for the degradation of the analytes. (A$_b$) and (B$_b$) depict degradation kinetics. (A$_c$) and (B$_c$) show the change in color before and after the reaction, with the structures of the analytes depicted in (A$_d$), (B$_d$). Panel A-B reprinted with permission from Bhosale *et al.* [257]. Copyright © 2016 WILEY-VCH Verlag GmbH & Co. KGaA, Weinheim. (C) Au flake catalytic activity toward methanol (D) comparison of flakes and commercial particles in the catalysis in panel. Panels (C-D) reprinted from Yang *et al.* [82].

Wenjing and his group explored the field of catalysis by harnessing the gold flakes as substrate electrode to construct bimetallic and trimetallic Au-based catalysts [224], with platinum (Pt) or palladium (Pd) deposited on the gold surface. These hybrid catalysts exhibited enhanced catalytic activity for methanol electrooxidation and demonstrated greater resistance to catalyst poisoning compared to commercial Pt-based catalysts. Similarly, methanol oxidation was obtained by Yang *et al.* [82] by employing a vapor-phase synthesis method to deposit the gold flakes onto fluorine-doped tin oxide (FTO) substrates, achieving significantly higher methanol oxidation rates compared to Au-deposited nanoparticles on the same FTO platform, as shown in **Figure 11**(C,D). Collectively, these studies illustrate how the unique properties of gold flakes—particularly their atomically flat surfaces and crystallographic orientation together with a large surface to volume ratio—enable improved catalytic performances. As they can be transferred to electrodes, electrocatalysis applications are a future research direction to be expected. A summary of reported studies integrating gold flakes for catalysis is provided in Table 5.

**Table 5: Gold flakes as a catalytic platform**

| Method | Application summary | Ref. |
|---|---|---|
| As synthesized | Graphene - Au flakes hybrid for high catalytic activity | [255]. |
| | Electrochemical catalytic oxidation of analytes by gold flakes on ITO | [222]. |
| | Catalytic degradation of Azo compounds using NaBH4 | [257]. |
| | Catalytic hydrogenation of furfural, and $H_2O_2$ catalytic decomposition | [237]. |
| | Catalytic electrooxidation of formic acid/methanol | [82, 223, 224]. |
| | Electrochemical catalytic activity of $H_2O_2$ | [225]. |

Note: "As synthesized" means gold flake without further fabrication or modification

## Other applications

Gold flakes have also found applications beyond the conventional realms of optics, sensing, and electronics. For example, Zhu *et al.* [258] utilized the atomically flat, highly reflective surface of gold flakes to assemble a miniature capacitive balance with a piconewton-level detection limit, suitable for measuring microscale object masses or weak forces (**Figure 12**(A)). In this design, the gold flake is suspended and serves as a mirror within a laser setup, achieving a detection limit as low as a few hundred nanograms.

Gold flakes have also shown remarkable biocompatibility. Singh *et al.* [259] investigated the interaction between gold flakes and mice peritoneal macrophages using live-cell confocal imaging and SEM (**Figure 12**(B)). Their study revealed that cells can selectively interact with gold flakes, either by internalizing them or exhibiting a frustrated phagocytosis-like phenomenon, particularly with larger gold flakes that cannot be internalized. This cellular interaction has been harnessed in drug delivery applications, where biocompatible gold flakes were loaded with drugs showing a sustained release, leading to antitumor activity in the cell line and suggesting their potential as novel drug carriers for breast cancer therapy. The tunable optical properties of large area flakes also enable their use as NIR-absorbing films or antennas, as reported by Shankar *et al.* [107] with a possible integration of the material into medical applications, e.g., hyperthermia treatment of cancer cells and in IR-absorbing optical coatings [45, 107]. Another biological application was reported by Radha *et al.* [260], who leveraged the biocompatibility and fluorescent signal enhancement properties of gold flakes for single-cell studies (**Figure 12**(C)). By positioning cells in proximity to the gold surface, they demonstrated an enhancement of fluorescence signals from fluorophores within the cells, offering exciting possibilities for probing the dynamic interactions between cells and their microenvironment.

On a smaller scale, a plasmonic optical antenna has successfully been designed to enable single-molecule biochemical investigations [261]. The antenna serves as an anchoring site for mechanically interlocked molecules. Gold flakes have also been introduced as versatile nanoscale building blocks. Ah *et al.* [262] reported the fabrication of nanocomponents and nanomachines—such as a nano wheel capable of moving by a few microns, fabricated using top-down strategies such as FIB and EBL (**Figure 12**(D)), exemplifying the convergence of "top-down" and "bottom-up" approaches in nanotechnology. Similarly, Yun *et al.* [263] fabricated nanoscale components from gold flakes - including nano blocks - that could be used in nanoelectromechanical systems, paving the way for the development of nano/micro-machines like nano wheels and nano saws. Also, in materials science the physical properties of gold flakes can be harnessed. Joo *et al.* [264] utilized the large atomically ordered surface area of the flake for heteroepitaxial deposition of continuous ZnO films through solution-based techniques (**Figure 12**(E)). This method facilitates detailed studies of metal-semiconductor junctions, photovoltaics, and new devices. Meanwhile, Sharma *et al.* [265] employed gold flakes as antennas for the optothermal assembly of colloidal silica microparticles near the flakes to explore the controlled, two-dimensional assembly of colloidal crystals through thermal gradients, contributing to the understanding of light-activated colloidal matter (Figure 12(F)).

The high precision of structures milled via Helium FIB from gold flakes also enables the realization of special optical antennas for efficient photon momentum transfer, used by Wu *et al.* [266] to create light-driven microdrones, later enhanced with embedded optical tweezers for nanoparticle manipulation [267] (**Figure 12**(G-H)).

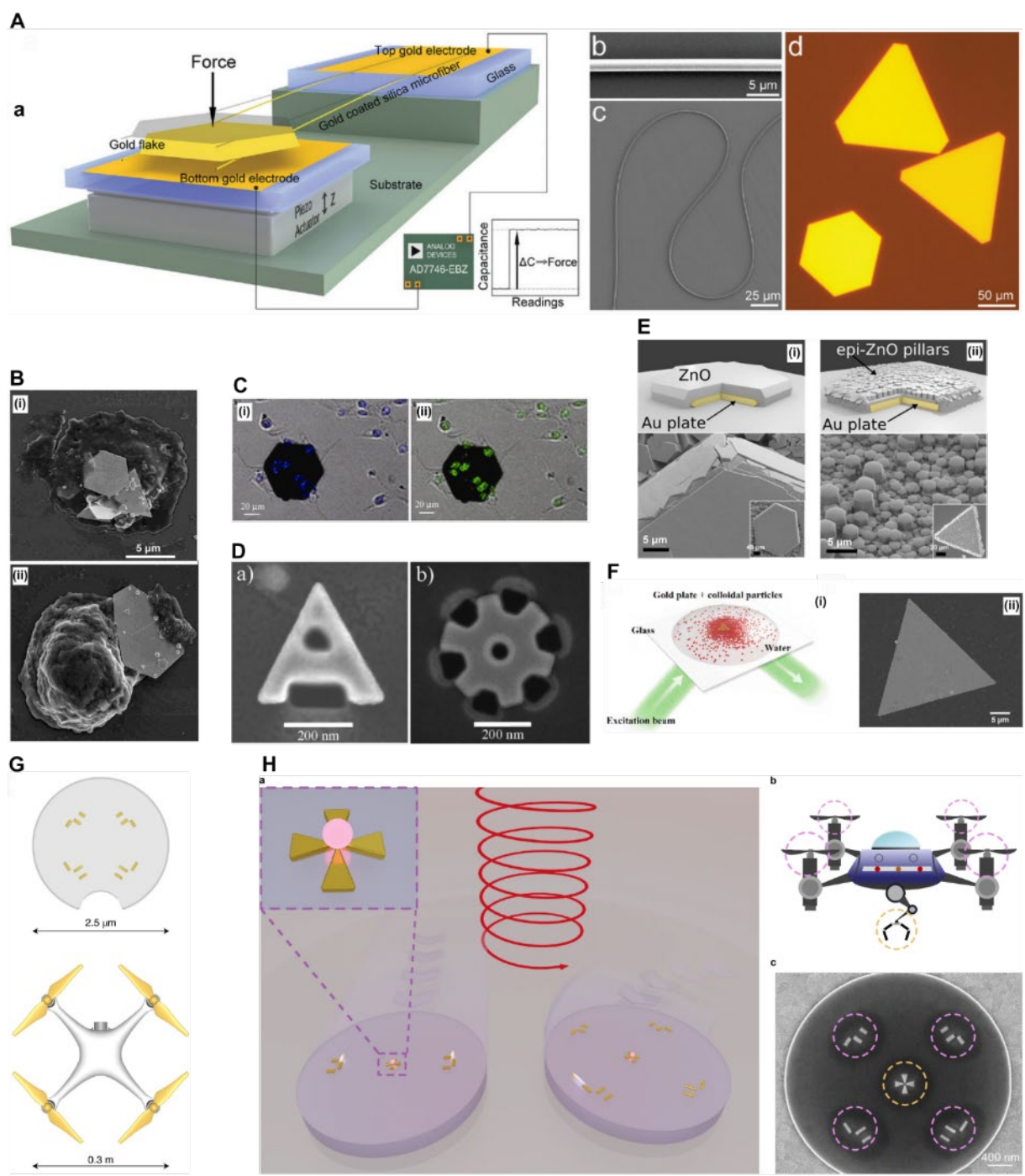

**Figure 12: Gold flakes in various fields.** (A$_a$) Schematic diagram of the miniature capacitive pico-balance. (A$_b$), (A$_c$) SEM images of straight and bent silica microfibers. (A$_d$) Optical micrograph of single-crystal gold flakes. Panel A reprinted with permission from Zhu *et al.* [258]. Copyright © 2024 Wiley-VCH GmbH. (B$_i$) SEM image of cells holding gold flake with membrane cup. (B$_{ii}$) Partially phagocytosed gold flake. Panel B reprinted (adapted) with permission from Singh *et al.* [259]. Copyright © 2014, American Chemical Society. (C$_i$), (C$_{ii}$) Confocal images of a gold flake hosting mouse 3T3 cells, stained with Hoechst and Alexa 488 to visualize the nucleus and the acetylated histone H3, respectively. Panel C adopted from Radha *et al.* [260]. Copyright © 2010, Tsinghua University Press and Springer-Verlag Berlin Heidelberg. (D$_a$), (D$_b$) Various nanocomponents, including nanogear, fabricated from gold flakes. Panel D reprinted (adapted) with permission from Ah *et al.* [262]. Copyright © 2005, American Chemical Society. (E$_i$) Schematic of flat ZnO on gold flake, showing the Au substrate beneath a smooth ZnO film, and an SEM micrograph of the smooth morphology of microwave-nucleated ZnO after growth. (E$_{ii}$) Nucleation and growth steps of oven-nucleated ZnO on gold, with an SEM micrograph showing rough, faceted morphology

after the growth. Panel E reprinted (adapted) with permission from Joo *et al.* [264]. Copyright © 2013, American Chemical Society. (F$_i$) Schematic illustration of colloidal particle assembly on and around a gold flake. The flake is represented as a yellow triangular plate, and the colloidal particles as red spheres initially dispersed in solution. Arrows indicate the direction of excitation. (F$_{ii}$) Field emission scanning electron microscopy (FESEM) image of a representative gold flake used in the experiments. Panel F was reused with permission from [265]. Copyright © 2020 IOP Publishing Ltd. (G) The top view of the four-motor microdrone (top) and comparison to a macroscopic quadcopter drone (bottom). Panel G reprinted (adapted) with permission from *Wu et al.* [266]. Copyright © 2022, The Author(s), under exclusive license to Springer Nature Limited. (H) Micro drones with embedded optical tweezers (H$_a$) Two types of micro drones, featuring either two or four nanoscale motors integrated with a tweezer-like manipulator, are optically actuated in an aqueous environment using an unfocused circularly polarized laser beam. Top left inset: An enlarged artistic illustration of a nanodiamond being trapped by the gold cross-antenna structure. (H$_b$) Conceptual schematic highlights the microrobot design, which integrates individually addressable motors for precise navigation and an independent gripping mechanism. (H$_c$) SEM image of the fabricated microrobot. The dashed circles in (H$_b$) and (H$_c$) correspond to structural features marked by the purple and yellow dashed circles in panel (H$_b$). Panel H reprinted from Qin *et al.* [267]. Copyright © 2025, The Author(s).

Finally, the thermal properties of gold flakes have also been explored in hybrid nanocomposite materials for tuning the photothermal properties, security printing applications, and even traditional drawing pigments [231, 268], further emphasizing the vast potential of this unique 2D material.

**Table 6:** Gold flakes in other applications

| Method | Application summary | Ref. |
|---|---|---|
| As synthesized | Gold flake surface as a miniature capacitive picobalances | [258]. |
| | Gold flake surface signal fluorescent enhancement for cell studies | [260]. |
| | Gold flake as a thermoplasmonic platform for assembly of colloids | [265]. |
| | Epitaxial Growth of ZnO on gold flake surface | [264]. |
| | Tuning the thermal properties of hybrid materials | [231]. |
| | Printing and even pigment application | [268]. |
| Structured | Plasmonic antenna to mechanically anchor organic molecules | [261]. |
| | Light-driven microdrones | [266, 267]. |
| | FIB/EBL based versatile nano-building blocks | [262, 263]. |
| Functionalized flakes | Gold flake for drug delivery and phagocytosis analysis | [259]. |

Note: "As synthesized" means gold flake without further fabrication or modification

## Further developments

Despite notable advancements in the field, such as reproducibility and high-yield synthesis of monodisperse, anisotropic gold nanostructures, milling methods, and reliable transfer of flakes or milled nanostructures between arbitrary surfaces [70, 115] translating these advances into scalable, cost-effective, and industrially viable fabrication strategies remains a central challenge. As the field moves forward, a key priority will be establishing cost-effective, sustainable strategies that bridge wet chemistry, bottom-up assembly, and CMOS-compatible top-down fabrication to enable the large-scale production of structurally defined gold nanostructure arrays.

Several promising directions have emerged to obtain a monocrystalline film through methods analogous to conventional gold evaporation or producing gold flake arrays via deterministic positioning in combined approaches. Capitaine *et al.* [269], for instance, demonstrated a hybrid approach combining bottom-up and top-down strategies, where assemblies of single-crystal gold nano cubes were epitaxially transformed into continuous monocrystalline plasmonic structures with arbitrary geometries. Epitaxial electrochemical deposition techniques have also enabled the growth of atomically flat, gold films with tunable thickness on top of a prefabricated monocrystalline silver surface [270]. Additionally, methods adapted from the classical silicon–germanium industry, such as the Czochralski process, have been explored for producing thin layers of monocrystalline gold. Vesseur *et al.* [129] demonstrated that this approach can yield suitable starting materials for photonic applications, provided that post-growth surface treatment is applied to restore atomic flatness.

In parallel, deterministic patterning strategies, such as seed-mediated growth on substrates pre-patterned via lithographic or vapor-phase processes, have allowed for the controlled formation of gold flake arrays with defined faceting and orientation [81, 271-274]. In particular, nanoimprint lithography combined with plasmon-mediated growth has yielded promising results in generating periodic arrays with high precision [271-273].

To overcome the remaining barriers in throughput and prototyping, alternative large-area patterning technologies with sub-15 nm resolution offer compelling advantages over conventional methods like FIB and EBL [275], accelerating the transition toward overcoming current limitations of rapid prototyping. To further address the challenges of cost and scalability, leveraging self-organization approaches is worth more attention to explore the controlled growth of functional structures and surface from the bottom up [276-279]. These include approaches where gold flakes are functionalized with DNA strands to enable programmable pattern formation [279], directed assembly into predefined geometries [278], or the formation of macroscopic hierarchical architectures from nanoscale building blocks [276, 277] and enriching the surface functionality [125, 280]. Altogether, the integration of these emerging strategies has the potential to unlock scalable, reproducible, and cost-effective platforms based on monocrystalline gold flakes. As these methods continue to mature, they are likely to establish gold flake architectures as a foundational material system for the next generation of universal, multifunctional devices. Finally, there are efforts to change the flakes themselves, as presented in [31] where during chemical thinning the crystal structure shows a phase transition from fcc to hcp at a thickness of ~12 nm. This leads to increased mechanical stability and promises even smaller devices to be developed in the future.

## Conclusions and Outlook

Monocrystalline gold flakes have established themselves as a uniquely versatile material system in micro and nano science, distinguished by their atomically flat surfaces, crystallographic precision, and large lateral dimensions confined to nanometric thickness. Their synthesis has evolved in the last two decades to reproducible high-yield methods—spanning chemical, physical, and bioinspired strategies, capable of producing flakes within a rather narrow size distribution, but across broad size scales. These features, combined with the inherent chemical stability of gold and its compatibility with various functionalization techniques, have positioned gold flakes as building blocks for diverse applications. Their integration into nanotechnology highlights the convergence of bottom-up synthesis and top-down methodology. These platforms show growing potential across diverse fields, including high-performance SERS substrates, tunable nanoantennas, and optical metasurfaces, to support catalytic systems, flexible electronic devices, and molecular electronics, gold flakes have demonstrated superior performance compared to conventional polycrystalline gold or evaporated films. Their utility extends to platforms requiring nanoscale precision and structural integrity, such as scanning probe microscopy, where their atomically smooth and conductive surfaces provide unmatched analytical advantages.

In the future, we expect more sophisticated growth protocols, including multi-step or regrowth methods, yielding gold flakes with millimeter-scale lateral dimensions yet retaining sub-100 nm thickness. These extended monocrystalline substrates will drive the development of novel low-loss, large area meta materials, integrated plasmonic circuits and highly reproducible sensor platforms. Another advancement will be the development of flakes with single nm thickness by new synthesis protocols or polishing. This will enhance surface effects like second harmonic generation or yield non-classical behavior due to the materials band structure becoming discretized – available than on large areas for novel meta-surface and plasmonic device functionalities.

The possibilities of electrically contacting plasmonic nanostructures are still to be explored and gold flakes will be indispensable in doing so. We expect novel sensing strategies based on surface potential dependent analyte response or new recipes to functionalize surface with selective receptor molecules. Also the atomically flat surface yields an opportunity to understand (electro-)catalytic processes better and better, but also to realize, e.g., nano-scaled photo-electrochemical laboratories, both integrated in lab-on-a-chip applications, as well as on the level of single molecules.

These prospects underscore the pivotal role of monocrystalline gold flakes in pushing the boundaries of nanoscale optics, sensing and establishing control over light-matter interaction on ever smaller scales, hopefully down to the atomic limit.

## Author information

**Amro Sweedan** – *The Ilse -Katz Institute for Nanoscale Science & Technology, Ben-Gurion University of the Negev, POB 653, Beer-Sheba Campus, Building 51, 8410501, Israel. Department of Solar Energy and Environmental Physics, Swiss Institute for Dryland Environmental and Energy Research, J. Blaustein Institutes for Desert Research, Ben-Gurion University of the Negev, Midreshset Ben-Gurion, Building 26, 8499000, Israel. Email: sweedan1@gmail.com; ORCID: https://orcid.org/0000-0001-9623-9726.*

**Kefan Zhang** – *Department of Solar Energy and Environmental Physics, Swiss Institute for Dryland Environmental and Energy Research, J. Blaustein Institutes for Desert Research, Ben-Gurion University of the Negev, Midreshset Ben-Gurion, Building 26, 8499000, Israel. Email: kefan@post.bgu.ac.il; ORCID: https://orcid.org/0000-0003-4019-4455.*

**Muhammad Y. Bashouti** – *Department of Solar Energy and Environmental Physics, Swiss Institute for Dryland Environmental and Energy Research, J. Blaustein Institutes for Desert Research, Ben-Gurion University of the Negev, Midreshset Ben-Gurion, Building 26, 8499000, Israel. The Ilse -Katz Institute for Nanoscale Science & Technology, Ben-Gurion University of the Negev, POB 653, Beer-Sheba Campus, Building 51, 8410501, Israel. Email: bashouti@bgu.ac.il; ORCID: https://orcid.org/0000-0002-0371-7088.*

**Thorsten Feichtner** – *Nano-Optics and Biophotonics Group, Experimental Physics 5, Institute of Physics, University of Würzburg, Am Hubland, 97074 Würzburg, Germany. Email:*

*thorsten.feichtner@physik.uni-wuerzburg.de;* ORCID: *https://orcid.org/0000-0002-0605-6481.*

## Note

The authors declare no conflict of interest.

## Bibliography

1.  Sun, Y. and Y. Xia, *Shape-Controlled Synthesis of Gold and Silver Nanoparticles.* Science, 2002. **298**(5601): p. 2176-2179.

2.  El-Sayed, M.A., *Some Interesting Properties of Metals Confined in Time and Nanometer Space of Different Shapes.* Accounts of Chemical Research, 2001. **34**(4): p. 257-264.

3.  Scarabelli, L., et al., *Plate-Like Colloidal Metal Nanoparticles*, in *Chemical Reviews*. 2023, American Chemical Society. p. 3493-3542.

4.  Tan, C. and H. Zhang, *Wet-chemical synthesis and applications of non-layer structured two-dimensional nanomaterials*, in *Nature Communications*. 2015.

5.  Nagpal, P., et al., *Ultrasmooth Patterned Metals for Plasmonics and Metamaterials.* Science, 2009. **325**(5940): p. 594-597.

6.  Ye, S., et al., *One-Step Preparation of Biocompatible Gold Nanoplates with Controlled Thickness and Adjustable Optical Properties for Plasmon-Based Applications.* Advanced Functional Materials, 2020. **30**(40).

7.  Liu, N., et al., *Nanoantenna-enhanced gas sensing in a single tailored nanofocus.* Nat Mater, 2011. **10**(8): p. 631-6.

8.  Huang, J.S., et al., *Atomically flat single-crystalline gold nanostructures for plasmonic nanocircuitry.* Nature Communications, 2010. **1**(9).

9.  Mahenderkar, N.K., et al., *Epitaxial lift-off of electrodeposited single-crystal gold foils for flexible electronics.* Science, 2017. **355**(6330): p. 1203-1206.

10. Moon, G.D., et al., *Highly stretchable patterned gold electrodes made of Au nanosheets.* Advanced Materials, 2013. **25**(19).

11. Maurer, J.H.M., et al., *Templated Self-Assembly of Ultrathin Gold Nanowires by Nanoimprinting for Transparent Flexible Electronics.* Nano Letters, 2016. **16**(5): p. 2921-2925.

12. Zhang, Y., et al., *A new preparation of Au nanoplates and their application for glucose sensing.* Biosensors and Bioelectronics, 2011. **28**(1): p. 344-348.

13.     Jin, Y., *Engineering plasmonic gold nanostructures and metamaterials for biosensing and nanomedicine.* Advanced Materials, 2012. **24**(38).

14.     Alivisatos, P., *The use of nanocrystals in biological detection*, in *Nature Biotechnology*. 2004.

15.     Yang, X., et al., *Gold Nanomaterials at Work in Biomedicine*, in *Chemical Reviews*. 2015.

16.     Bi, Y. and G. Lu, *Morphological controlled synthesis and catalytic activities of gold nanocrystals.* Materials Letters, 2008. **62**(17-18).

17.     Astruc, D., F. Lu, and J.R. Aranzaes, *Nanoparticles as Recyclable Catalysts: The Frontier between Homogeneous and Heterogeneous Catalysis.* Angewandte Chemie International Edition, 2005. **44**(48): p. 7852-7872.

18.     Dreaden, E.C., et al., *The golden age: gold nanoparticles for biomedicine.* Chemical Society Reviews, 2012. **41**(7): p. 2740-2779.

19.     Scarabelli, L., et al., *Monodisperse gold nanotriangles: size control, large-scale self-assembly, and performance in Surface-Enhanced Raman Scattering.* Acs Nano, 2014. **8**(6): p. 5833-5842.

20.     Saha, K., et al., *Gold Nanoparticles in Chemical and Biological Sensing.* Chemical Reviews, 2012. **112**(5): p. 2739-2779.

21.     Elghanian, R., et al., *Selective Colorimetric Detection of Polynucleotides Based on the Distance-Dependent Optical Properties of Gold Nanoparticles.* Science, 1997. **277**(5329): p. 1078-1081.

22.     Kim, F., J.H. Song, and P. Yang, *Photochemical Synthesis of Gold Nanorods.* Journal of the American Chemical Society, 2002. **124**(48): p. 14316-14317.

23.     Yu, et al., *Gold Nanorods: Electrochemical Synthesis and Optical Properties.* The Journal of Physical Chemistry B, 1997. **101**(34): p. 6661-6664.

24.     Jin, R., et al., *Controlling anisotropic nanoparticle growth through plasmon excitation.* Nature, 2003. **425**(6957): p. 487-490.

25.     Viarbitskaya, S., et al., *Tailoring and imaging the plasmonic local density of states in crystalline nanoprisms.* Nature Materials, 2013. **12**(5): p. 426-432.

26.     Qin, L., et al., *Nanodisk Codes.* Nano Letters, 2007. **7**(12): p. 3849-3853.

27.     Huang, X., et al., *Photochemically Controlled Synthesis of Anisotropic Au Nanostructures: Platelet-like Au Nanorods and Six-Star Au Nanoparticles.* ACS Nano, 2010. **4**(10): p. 6196-6202.

28.     Gong, S., et al., *A wearable and highly sensitive pressure sensor with ultrathin gold nanowires.* Nature Communications, 2014. **5**(1): p. 3132.

29.     Jana, N.R., L. Gearheart, and C.J. Murphy, *Seed-Mediated Growth Approach for Shape-Controlled Synthesis of Spheroidal and Rod-like Gold Nanoparticles Using a Surfactant Template.* Advanced Materials, 2001. **13**(18): p. 1389-1393.

30.     Millstone, J.E., et al., *Colloidal Gold and Silver Triangular Nanoprisms.* Small, 2009. **5**(6): p. 646-664.

31.     Zhang, T., et al., *Challenging the ideal strength limit in single-crystalline gold nanoflakes through phase engineering.* Nature Communications, 2025. **16**(1): p. 926.

32.     Tao, A.R., S. Habas, and P. Yang, *Shape control of colloidal metal nanocrystals*, in *Small*. 2008.

33.     Sun, X., S. Dong, and E. Wang, *Large-scale, Solution-phase Production of Microsized, Single-crystalline, Hexagonal Gold Microplates by Thermal Reduction of HAuCl4 with Oxalic Acid.* Chemistry Letters, 2005. **34**(7): p. 968-969.

34.     Qin, H.L., et al., *Thickness-controlled synthesis of ultrathin au sheets and surface plasmonic property.* Journal of the American Chemical Society, 2013. **135**(34).

35.     Sweedan, A.O., et al., *Evolutionary Optimized, Monocrystalline Gold Double Wire Gratings as a Novel SERS Sensing Platform.* Small, 2024. **20**(35).

36.     Pan, C., et al., *Large area single crystal gold of single nanometer thickness for nanophotonics.* Nature Communications, 2024. **15**(1).

37.     Turkevich, J., P.C. Stevenson, and J. Hillier, *A study of the nucleation and growth processes in the synthesis of colloidal gold.* Discussions of the Faraday Society, 1951. **11**(0): p. 55-75.

38.     Brüche, B., *Über sehr dünne Goldeinkristall-Plättchen.* Kolloid-Zeitschrift, 1960. **170**(2).

39.     MORISS, R.H., R.G. PEACOCK, and W.O. MILLIGAN, *Dislocations in Lamellar Crystals of Colloidal Gold.* Journal of Electron Microscopy, 1965. **14**(2): p. 93-100.

40.     Phys, et al., *Preparation of Electron-Transparent (111) Gold Platelets as Substrates for Epitaxial Studies*, in *Advan. Electron. Electron Phys*. 1969. p. 698-698.

41.     Kirkland, A.I., et al., *Structural studies of trigonal lamellar particles of gold and silver.* Proceedings of the Royal Society of London Series a-Mathematical Physical and Engineering Sciences, 1993. **440**(1910): p. 589-609.

42.     Zhou, J., et al., *Macroscopic Single-Crystal Gold Microflakes and Their Devices.* Advanced Materials, 2015. **27**(11): p. 1945-1950.

43.     Jin, R., et al., *Photoinduced conversion of silver nanospheres to nanoprisms.* Science, 2001. **294**(5548).

44.     Kim, J.U., et al., *Preparation of Gold Nanowires and Nanosheets in Bulk Block Copolymer Phases under Mild Conditions.* Advanced Materials, 2004. **16**(5).

45.    Shankar, S.S., et al., *Biological synthesis of triangular gold nanoprisms.* Nature Materials, 2004. **3**(7): p. 482-488.

46.    Sun, X., S. Dong, and E. Wang, *Large-scale synthesis of micrometer-scale single-crystalline Au plates of nanometer thickness by a wet-chemical route.* Angewandte Chemie - International Edition, 2004. **43**(46): p. 6360-6363.

47.    Cui, X., et al., *Circular Gold Nanodisks with Synthetically Tunable Diameters and Thicknesses.* Advanced Functional Materials, 2018. **28**(11): p. 1705516.

48.    Park, J.H., et al., *Control of Electron Beam-Induced Au Nanocrystal Growth Kinetics through Solution Chemistry.* Nano Letters, 2015. **15**(8).

49.    Ibano, D., Y. Yokota, and T. Tominaga, *Preparation of Gold Nanoplates Protected by an Anionic Phospholipid.* Chemistry Letters, 2003. **32**(7): p. 574-575.

50.    Ortiz-Castillo, J.E., et al., *Anisotropic gold nanoparticles: A survey of recent synthetic methodologies.* Coordination Chemistry Reviews, 2020. **425**: p. 213489.

51.    Zheng, J., et al., *Gold Nanorods: The Most Versatile Plasmonic Nanoparticles*, in *Chemical Reviews*. 2021.

52.    Li, N., P. Zhao, and D. Astruc, *Anisotropic gold nanoparticles: Synthesis, properties, applications, and toxicity*, in *Angewandte Chemie - International Edition*. 2014.

53.    Lv, L., et al., *Trans-Scale 2D Synthesis of Millimeter-Large Au Single Crystals via Silk Fibroin Templates.* ACS Sustainable Chemistry & Engineering, 2018. **6**(9): p. 12419-12425.

54.    Millstone, J.E., et al., *Observation of a Quadrupole Plasmon Mode for a Colloidal Solution of Gold Nanoprisms.* Journal of the American Chemical Society, 2005. **127**(15): p. 5312-5313.

55.    Kelly, K.L., et al., *The Optical Properties of Metal Nanoparticles: The Influence of Size, Shape, and Dielectric Environment.* The Journal of Physical Chemistry B, 2003. **107**(3): p. 668-677.

56.    Shuford, K.L., M.A. Ratner, and G.C. Schatz, *Multipolar excitation in triangular nanoprisms.* The Journal of Chemical Physics, 2005. **123**(11): p. 114713.

57.    Kiani, F. and G. Tagliabue, *High Aspect Ratio Au Microflakes via Gap-Assisted Synthesis.* Chemistry of Materials, 2022. **34**(3): p. 1278-1288.

58.    Millstone, J.E., G.S. Métraux, and C.A. Mirkin, *Controlling the Edge Length of Gold Nanoprisms via a Seed-Mediated Approach.* Advanced Functional Materials, 2006. **16**(9): p. 1209-1214.

59.    Hoffmann, B., et al., *New insights into colloidal gold flakes: structural investigation, micro-ellipsometry and thinning procedure towards ultrathin monocrystalline layers.* Nanoscale, 2016. **8**(8): p. 4529-4536.

60. Natarajan, P., et al., *The non-stationary case of the Maxwell-Garnett theory: growth of nanomaterials (2D gold flakes) in solution.* Nanoscale Advances, 2020. **2**(3): p. 1066-1073.

61. Lofton, C. and W. Sigmund, *Mechanisms Controlling Crystal Habits of Gold and Silver Colloids.* Advanced Functional Materials, 2005. **15**(7): p. 1197-1208.

62. Alloyeau, D., et al., *Unravelling Kinetic and Thermodynamic Effects on the Growth of Gold Nanoplates by Liquid Transmission Electron Microscopy.* Nano Letters, 2015. **15**(4): p. 2574-2581.

63. Elechiguerra, J.L., J. Reyes-Gasga, and M.J. Yacaman, *The role of twinning in shape evolution of anisotropic noble metal nanostructures.* Journal of Materials Chemistry, 2006. **16**(40): p. 3906-3919.

64. *Flake growth simulation.* Available from: https://github.com/Rene-007/flake_growth.

65. Boroviks, S., et al., *Interference in edge-scattering from monocrystalline gold flakes [Invited].* Optical Materials Express, 2018. **8**(12): p. 3688-3697.

66. LaMer, V.K. and R.H. Dinegar, *Theory, Production and Mechanism of Formation of Monodispersed Hydrosols.* Journal of the American Chemical Society, 1950. **72**(11): p. 4847-4854.

67. Van Hove, M.A., et al., *The surface reconstructions of the (100) crystal faces of iridium, platinum and gold: I. Experimental observations and possible structural models.* Surface Science, 1981. **103**(1): p. 189-217.

68. Xia, Y., et al., *Shape-controlled synthesis of metal nanocrystals: Simple chemistry meets complex physics?*, in *Angewandte Chemie - International Edition.* 2009. p. 60-103.

69. Huang, J., et al., *Bio-inspired synthesis of metal nanomaterials and applications.* Chemical Society Reviews, 2015. **44**(17): p. 6330-6374.

70. Krauss, E., et al., *Controlled growth of high-aspect-ratio single-crystalline gold platelets.* Crystal Growth & Design, 2018. **18**(3): p. 1297-1302.

71. Grace, A.N. and K. Pandian, *One pot synthesis of polymer protected gold nanoparticles and nanoprisms in glycerol.* Colloids and Surfaces A: Physicochemical and Engineering Aspects, 2006. **290**(1): p. 138-142.

72. Lohse, S.E., et al., *Anisotropic Noble Metal Nanocrystal Growth: The Role of Halides.* Chemistry of Materials, 2014. **26**(1): p. 34-43.

73. Ha, T.H., H.-J. Koo, and B.H. Chung, *Shape-Controlled Syntheses of Gold Nanoprisms and Nanorods Influenced by Specific Adsorption of Halide Ions.* The Journal of Physical Chemistry C, 2007. **111**(3): p. 1123-1130.

74. Hu, H., et al., *Two-Dimensional Au Nanocrystals: Shape/Size Controlling Synthesis, Morphologies, and Applications*, in *Particle and Particle Systems Characterization.* 2015, Wiley-VCH Verlag. p. 796-808.

75.     Gao, X., et al., *Zwitterionic vesicles with AuCl4− counterions as soft templates for the synthesis of gold nanoplates and nanospheres.* Chemical Communications, 2014. **50**(63): p. 8783-8786.

76.     Li, C., et al., *Mass synthesis of large, single-crystal Au nanosheets based on a polyol process.* Advanced Functional Materials, 2006. **16**(1): p. 83-90.

77.     Großmann, S., et al., *Nonclassical Optical Properties of Mesoscopic Gold.* Physical Review Letters, 2019. **122**(24): p. 246802.

78.     Xia, Y., et al., *Seed-Mediated Growth of Colloidal Metal Nanocrystals*, in *Angewandte Chemie - International Edition*. 2017.

79.     Zhu, J., et al., *Synthesis and growth mechanism of gold nanoplates with novel shapes.* Journal of Crystal Growth, 2011. **321**(1).

80.     Bolisetty, S., et al., *Amyloid-mediated synthesis of giant, fluorescent, gold single crystals and their hybrid sandwiched composites driven by liquid crystalline interactions.* Journal of Colloid and Interface Science, 2011. **361**(1).

81.     Demille, T.B., et al., *Epitaxially aligned single-crystal gold nanoplates formed in large-area arrays at high yield.* Nano Research, 2022. **15**(1): p. 296-303.

82.     Yang, S., et al., *Low-Temperature Vapor-Phase Synthesis of Single-Crystalline Gold Nanostructures: Toward Exceptional Electrocatalytic Activity for Methanol Oxidation Reaction.* Nanomaterials, 2019. **9**(4): p. 595.

83.     Xia, Y., X. Xia, and H.-C. Peng, *Shape-Controlled Synthesis of Colloidal Metal Nanocrystals: Thermodynamic versus Kinetic Products.* Journal of the American Chemical Society, 2015. **137**(25): p. 7947-7966.

84.     Murphy, C.J., et al., *Gold Nanoparticles in Biology: Beyond Toxicity to Cellular Imaging.* Accounts of Chemical Research, 2008. **41**(12): p. 1721-1730.

85.     Chen, S., et al., *Rapid Seedless Synthesis of Gold Nanoplates with Microscaled Edge Length in a High Yield and Their Application in SERS.* Nano-Micro Letters, 2016. **8**(4): p. 328-335.

86.     Wu, X., et al., *Single-crystalline gold microplates grown on substrates by solution-phase synthesis.* Crystal Research and Technology, 2015. **50**(8): p. 595-602.

87.     Xin, W., et al., *One-Step Synthesis of Tunable-Size Gold Nanoplates on Graphene Multilayers.* Nano Letters, 2018. **18**(3).

88.     Niu, J., et al., *Novel polymer-free iridescent lamellar hydrogel for two-dimensional confined growth of ultrathin gold membranes.* Nature Communications, 2014. **5**(1): p. 3313.

89.     Zhang, J., et al., *Shape-selective synthesis of gold nanoparticles with controlled sizes, shapes, and plasmon resonances.* Advanced Functional Materials, 2007. **17**(16).

90. Farkas, S., et al., *Reaction-Diffusion Assisted Synthesis of Gold Nanoparticles: Route from the Spherical Nano-Sized Particles to Micrometer-Sized Plates.* Journal of Physical Chemistry C, 2021. **125**(47): p. 26116-26124.

91. Sau, T.K. and C.J. Murphy, *Room Temperature, High-Yield Synthesis of Multiple Shapes of Gold Nanoparticles in Aqueous Solution.* Journal of the American Chemical Society, 2004. **126**(28): p. 8648-8649.

92. Ali Umar, A., et al., *Formation of highly thin, electron-transparent gold nanoplates from nanoseeds in ternary mixtures of cetyltrimethylammonium bromide, poly(vinyl pyrrolidone), and poly(ethylene glycol).* Crystal Growth and Design, 2010. **10**(8).

93. Wang, G., et al., *High-yield halide-free synthesis of biocompatible Au nanoplates.* Chemical Communications, 2016. **52**(2): p. 398-401.

94. Wang, Z., et al., *Synthesis, characterization and mechanism of cetyltrimethylammonium bromide bilayer-encapsulated gold nanosheets and nanocrystals.* Applied Surface Science, 2008. **254**(20): p. 6289-6293.

95. Wang, L.Y., et al., *Controlled formation of gold nanoplates and nanobelts in lyotropic liquid crystal phases with imidazolium cations.* Colloids and Surfaces A: Physicochemical and Engineering Aspects, 2007. **293**(1): p. 95-100.

96. Umar, A.A., et al., *Formation of high-yield gold nanoplates on the surface: Effective two-dimensional crystal growth of nanoseed in the presence of poly(vinylpyrrolidone) and cetyltrimethylammonium bromide.* Crystal Growth and Design, 2009. **9**(6).

97. Fievet, F., J.P. Lagier, and M. Figlarz, *Preparing Monodisperse Metal Powders in Micrometer and Submicrometer Sizes by the Polyol Process.* MRS Bulletin, 1989. **14**(12).

98. Koczkur, K.M., et al., *Polyvinylpyrrolidone (PVP) in nanoparticle synthesis.* Dalton Transactions, 2015. **44**(41): p. 17883-17905.

99. Sun, X. and Y. Luo, *Synthesis of gold microplates and polyhedral nanoparticles.* Materials Letters, 2006. **60**(24).

100. Liu, X., et al., *Shape-controlled growth of micrometer-sized gold crystals by a slow reduction method.* Small, 2006. **2**(8-9).

101. Wang, L., et al., *Synthesis of Gold Nano- and Microplates in Hexagonal Liquid Crystals.* The Journal of Physical Chemistry B, 2005. **109**(8): p. 3189-3194.

102. Xie, J., et al., *Identification of active biomolecules in the high-yield synthesis of single-crystalline gold nanoplates in algal solutions.* Small, 2007. **3**(4).

103. Shao, Y., Y. Jin, and S. Dong, *Synthesis of gold nanoplates by aspartate reduction of gold chloride.* Chemical Communications, 2004. **4**(9): p. 1104-1105.

104. Tan, Y.N., J.Y. Lee, and D.I.C. Wang, *Aspartic acid synthesis of crystalline gold nanoplates, nanoribbons, and nanowires in aqueous solutions.* Journal of Physical Chemistry C, 2008. **112**(14).

105.    Wang, X., et al., *Vapor-phase preparation of single-crystalline thin gold microplates using HAuCl4 as the precursor for plasmonic applications.* RSC Advances, 2016. **6**(78): p. 74937-74943.

106.    Grzelczak, M., et al., *Shape control in gold nanoparticle synthesis.* Chemical Society Reviews, 2008. **37**(9): p. 1783-1791.

107.    Shankar, S.S., et al., *Controlling the optical properties of lemongrass extract synthesized gold nanotriangles and potential application in infrared-absorbing optical coatings.* Chemistry of Materials, 2005. **17**(3).

108.    Ankamwar, B., M. Chaudhary, and M. Sastry, *Gold nanotriangles biologically synthesized using tamarind leaf extract and potential application in vapor sensing.* Synthesis and Reactivity in Inorganic, Metal-Organic and Nano-Metal Chemistry, 2005. **35**(1).

109.    Alex, S., et al., *Simple and rapid green synthesis of micrometer scale single crystalline gold nanoplates using chitosan as the reducing agent.* Journal of Crystal Growth, 2014. **406**.

110.    Wei, D., et al., *Mass synthesis of single-crystal gold nanosheets based on chitosan.* Carbohydrate Research, 2007. **342**(16).

111.    Xie, J., et al., *High-Yield Synthesis of Complex Gold Nanostructures in a Fungal System.* The Journal of Physical Chemistry C, 2007. **111**(45): p. 16858-16865.

112.    Yoo, Y., et al., *Surfactant-Free Vapor-Phase Synthesis of Single-Crystalline Gold Nanoplates for Optimally Bioactive Surfaces.* Chemistry of Materials, 2017. **29**(20).

113.    Huang, S., et al., *Electrochemical synthesis of gold nanocrystals and their 1D and 2D organization.* Journal of Physical Chemistry B, 2005. **109**(42).

114.    Kou, J., C. Bennett-Stamper, and R.S. Varma, *Green Synthesis of Noble Nanometals (Au, Pt, Pd) Using Glycerol under Microwave Irradiation Conditions.* ACS Sustainable Chemistry & Engineering, 2013. **1**(7): p. 810-816.

115.    Wu, X.F., et al., *Silica-gold bilayer-based transfer of focused ion beam-fabricated nanostructures.* Nanoscale, 2015. **7**(39): p. 16427-16433.

116.    Lee, Y.-J., et al., *Ultrasmooth, Highly Spherical Monocrystalline Gold Particles for Precision Plasmonics.* ACS Nano, 2013. **7**(12): p. 11064-11070.

117.    Staechelin, Y.U., et al., *Size-Dependent Electron–Phonon Coupling in Monocrystalline Gold Nanoparticles.* ACS Photonics, 2021. **8**(3): p. 752-757.

118.    Chow, T.H., et al., *Gold Nanobipyramids: An Emerging and Versatile Type of Plasmonic Nanoparticles.* Accounts of Chemical Research, 2019. **52**(8): p. 2136-2146.

119.    Kullock, R., et al., *Electrically-driven Yagi-Uda antennas for light.* Nature Communications, 2020. **11**(1).

120.    Boroviks, S., et al., *Use of monocrystalline gold flakes for gap plasmon-based metasurfaces operating in the visible.* Optical Materials Express, 2019. **9**(11): p. 4209-4209.

121.    Méjard, R., et al., *Advanced engineering of single-crystal gold nanoantennas.* Optical Materials Express, 2017. **7**(4): p. 1157-1157.

122.    Méjard, R., et al., *Energy-Resolved Hot-Carrier Relaxation Dynamics in Monocrystalline Plasmonic Nanoantennas.* ACS Photonics, 2016. **3**(8).

123.    Chen, K., et al., *Electromechanically Tunable Suspended Optical Nanoantenna.* Nano Letters, 2016. **16**(4): p. 2680-2685.

124.    Song, M., et al., *Photoluminescence plasmonic enhancement of single quantum dots coupled to gold microplates.* Journal of Physical Chemistry C, 2014. **118**(16).

125.    Love, J.C., et al., *Self-assembled monolayers of thiolates on metals as a form of nanotechnology*, in *Chemical Reviews*. 2005.

126.    Baigorri, R., et al., *Optical enhancing properties of anisotropic gold nanoplates prepared with different fractions of a natural humic substance.* Chemistry of Materials, 2008. **20**(4).

127.    Faraday, M., *X. The Bakerian Lecture. —Experimental relations of gold (and other metals) to light.* 1857. **147**: p. 145-181.

128.    Ashcroft, N.W. and N.D. Mermin, *Solid State Physics*. 1976: Holt-Saunders.

129.    Vesseur, E.J.R., et al., *Surface plasmon polariton modes in a single-crystal Au nanoresonator fabricated using focused-ion-beam milling.* Applied Physics Letters, 2008. **92**(8).

130.    Maier, S.A., *Plasmonics: Fundamentals and Applications*. 2007: Springer New York, NY.

131.    Frank, B., et al., *Short-range surface plasmonics: Localized electron emission dynamics from a 60-nm spot on an atomically flat single-crystalline gold surface.* Science Advances, 2017. **3**(7): p. e1700721.

132.    Kaltenecker, K.J., et al., *Mono-crystalline gold platelets: A high-quality platform for surface plasmon polaritons.* Nanophotonics, 2020. **9**(2): p. 509-522.

133.    Boroviks, S., et al., *Extremely confined gap plasmon modes: when nonlocality matters.* Nature Communications, 2022. **13**(1): p. 3105.

134.    Wang, J., et al., *Strong vibrational coupling in room temperature plasmonic resonators.* Nature Communications, 2019. **10**(1).

135.    Wu, C.Y., et al., *Plasmonic green nanolaser based on a metal-oxide-semiconductor structure.* Nano Letters, 2011. **11**(10): p. 4256-4260.

136. Davis, T.J., et al., *Ultrafast vector imaging of plasmonic skyrmion dynamics with deep subwavelength resolution.* Science, 2020. **368**(6489): p. eaba6415.

137. Bowman, A.R., et al., *Quantum-mechanical effects in photoluminescence from thin crystalline gold films.* Light: Science and Applications, 2024. **13**(1).

138. Wu, C., et al., *Recrystallization-Enabled Fabrication of Single-Crystalline Gold Flakes for Plasmonic Applications.* ACS Applied Materials & Interfaces, 2024. **16**(49): p. 68204-68210.

139. Wiley, B.J., et al., *Fabrication of Surface Plasmon Resonators by Nanoskiving Single-Crystalline Gold Microplates.* Nano Letters, 2008. **8**(9): p. 3023-3028.

140. Novotny, L., *Effective Wavelength Scaling for Optical Antennas.* Physical Review Letters, 2007. **98**(26): p. 266802.

141. Biagioni, P., J.S. Huang, and B. Hecht, *Nanoantennas for visible and infrared radiation.* Reports on Progress in Physics, 2012. **75**(2).

142. Chen, K., et al., *High-Q, low-mode-volume and multiresonant plasmonic nanoslit cavities fabricated by helium ion milling.* Nanoscale, 2018. **10**(36).

143. Meier, J., et al., *Controlling Field Asymmetry in Nanoscale Gaps for Second Harmonic Generation.* Advanced Optical Materials, 2023. **11**(21): p. 2300731.

144. Liu, L., et al., *Atomically Smooth Single-Crystalline Platform for Low-Loss Plasmonic Nanocavities.* Nano Letters, 2022. **22**(4): p. 1786-1794.

145. Prangsma, J.C., et al., *Electrically Connected Resonant Optical Antennas.* Nano Letters, 2012. **12**(8): p. 3915-3919.

146. Kern, J., et al., *Electrically driven optical antennas.* Nature Photonics, 2015. **9**(9): p. 582-586.

147. Peng, W., et al., *Construction of nanoparticle-on-mirror nanocavities and their applications in plasmon-enhanced spectroscopy.* Chemical Science, 2024. **15**(8): p. 2697-2711.

148. Yan, Z., et al., *Enhancing Nanoparticle Electrodynamics with Gold Nanoplate Mirrors.* Nano Letters, 2014. **14**(5): p. 2436-2442.

149. Kumar, U., et al., *Single plasmon spatial and spectral sorting on a crystalline two-dimensional plasmonic platform.* Nanoscale, 2020. **12**(25): p. 13414-13420.

150. Kumar, S., et al., *Efficient Coupling of Single Organic Molecules to Channel Plasmon Polaritons Supported by V-Grooves in Monocrystalline Gold.* ACS Photonics, 2020. **7**(8): p. 2211-2218.

151. Siampour, H., et al., *Ultrabright single-photon emission from germanium-vacancy zero-phonon lines: deterministic emitter-waveguide interfacing at plasmonic hot spots.* 2020. **9**(4): p. 953-962.

152. Vasilii, V.K., M. Ducloy, and V.S. Letokhov, *Spontaneous emission of an atom in the presence of nanobodies.* Quantum Electronics, 2001. **31**(7): p. 569.

153. Purcell, E.M., *Spontaneous Emission Probabilities at Radio Frequencies.* Physical Review, 1946. **69**(11-12): p. 681.

154. Azzam, S.I., et al., *Ten years of spasers and plasmonic nanolasers.* Light: Science & Applications, 2020. **9**(1): p. 90.

155. Boroviks, S. and O.J.F. Martin, *Monocrystalline Gold Metasurface to Control Anisotropic Second-Harmonic Generation.* Advanced Optical Materials, 2025.

156. Boroviks, S., et al., *Anisotropic second-harmonic generation from monocrystalline gold flakes.* Optics Letters, 2021. **46**(4): p. 833-836.

157. Rodríguez Echarri, Á., et al., *Nonlinear Photoluminescence in Gold Thin Films.* ACS Photonics, 2023. **10**(8): p. 2918-2929.

158. Menabde, S.G., et al., *Near-field probing of image phonon-polaritons in hexagonal boron nitride on gold crystals.* Science Advances, 2022. **8**(28): p. eabn0627.

159. Geisler, P., et al., *Transmission of Plasmons through a Nanowire.* ACS Photonics, 2017. **4**(7).

160. Bordo, V.G., *Model of Fabry-P\'erot-type electromagnetic modes of a cylindrical nanowire.* Physical Review B, 2010. **81**(3): p. 035420.

161. Geisler, P., et al., *Multimode Plasmon Excitation and In Situ Analysis in Top-Down Fabricated Nanocircuits.* Physical Review Letters, 2013. **111**(18): p. 183901.

162. Krauss, E., et al., *Reversible Mapping and Sorting the Spin of Photons on the Nanoscale: A Spin-Optical Nanodevice.* Nano Letters, 2019. **19**(5): p. 3364-3369.

163. Schurr, B., et al., *Plasmonic Su-Schrieffer-Heeger chains with strong coupling amplitudes.* arXiv preprint arXiv:2504.02603, 2025.

164. Pres, S., et al., *Detection of a plasmon-polariton quantum wave packet.* Nature Physics, 2023. **19**(5): p. 656-662.

165. Krauss, E., et al., *Reversible Mapping and Sorting the Spin of Photons on the Nanoscale: A Spin-Optical Nanodevice.* Nano Letters, 2019. **19**(5): p. 3364-3369.

166. Major, T.A., et al., *Optical and Dynamical Properties of Chemically Synthesized Gold Nanoplates.* The Journal of Physical Chemistry C, 2013. **117**(3): p. 1447-1452.

167. Karaman, C.O., et al., *Ultrafast hot-carrier dynamics in ultrathin monocrystalline gold.* Nature Communications, 2024. **15**(1): p. 703.

168. Lebsir, Y., et al., *Ultimate Limit for Optical Losses in Gold, Revealed by Quantitative Near-Field Microscopy.* Nano Letters, 2022. **22**(14): p. 5759-5764.

169. Goetz, S., et al., *Investigation of the nonlinear refractive index of single-crystalline thin gold films and plasmonic nanostructures.* Applied Physics B, 2016. **122**(4): p. 94.

170. Cuche, A., et al., *Modal engineering of Surface Plasmons in apertured Au Nanoprisms.* Scientific Reports, 2015. **5**.

171. Kaltenecker, K.J., et al., *Near-infrared nanospectroscopy using a low-noise supercontinuum source.* APL Photonics, 2021. **6**(6).

172. Abbasirad, N., et al., *Near-field interference map due to a dipolar emission near the edge of a monocrystalline gold platelet.* Journal of Optics, 2022. **24**(12): p. 125001.

173. Abbasirad, N., et al., *Near-field launching and mapping unidirectional surface plasmon polaritons using an automated dual-tip scanning near-field optical microscope.* Photonics Research, 2022. **10**(11): p. 2628-2641.

174. Spektor, G., et al., *Revealing the subfemtosecond dynamics of orbital angular momentum in nanoplasmonic vortices.* Science, 2017. **355**(6330): p. 1187-1191.

175. Lin, Z.-H., et al., *Impact of Plasmonic and Dielectric Substrates on the Whispering-Gallery Modes in Self-Assembled Fluorescent Semiconductor Polymer Microspheres.* Nano Letters, 2023. **23**(14): p. 6512-6519.

176. Chen, Y., et al. *Optical Control of the Localized Surface Plasmon Resonance in a Heterotype and Hollow Gold Nanosheet.* Nanomaterials, 2023. **13**, DOI: 10.3390/nano13121826.

177. Viarbitskaya, S., et al., *Plasmonic Hot Printing in Gold Nanoprisms.* ACS Photonics, 2015. **2**(6): p. 744-751.

178. Celebrano, M., et al., *Mode matching in multiresonant plasmonic nanoantennas for enhanced second harmonic generation.* Nature Nanotechnology, 2015. **10**(5): p. 412-417.

179. Chen, W.-L., et al., *The Modulation Effect of Transverse, Antibonding, and Higher-Order Longitudinal Modes on the Two-Photon Photoluminescence of Gold Plasmonic Nanoantennas.* ACS Nano, 2014. **8**(9): p. 9053-9062.

180. Klaer, P., et al., *Polarization dependence of plasmonic near-field enhanced photoemission from cross antennas.* Applied Physics B, 2016. **122**(5): p. 136.

181. Lin, D. and J.-S. Huang, *Slant-gap plasmonic nanoantennas for optical chirality engineering and circular dichroism enhancement.* Optics Express, 2014. **22**(7): p. 7434-7445.

182. Celebrano, M., et al., *Evidence of Cascaded Third-Harmonic Generation in Noncentrosymmetric Gold Nanoantennas.* Nano Letters, 2019. **19**(10): p. 7013-7020.

183. Di Francescantonio, A., et al., *Coherent Control of the Nonlinear Emission of Single Plasmonic Nanoantennas by Dual-Beam Pumping.* Advanced Optical Materials, 2022. **10**(20): p. 2200757.

184. Huang, J.-S., et al., *Mode Imaging and Selection in Strongly Coupled Nanoantennas.* Nano Letters, 2010. **10**(6): p. 2105-2110.

185. Klaer, P., et al., *Robustness of plasmonic angular momentum confinement in cross resonant optical antennas.* Applied Physics Letters, 2015. **106**(26).

186. Liu, T.M., et al., *Nanoscale Confinement of All-Optical Magnetic Switching in TbFeCo - Competition with Nanoscale Heterogeneity.* Nano Letters, 2015. **15**(10).

187. Kejík, L., et al., *Structural and optical properties of monocrystalline and polycrystalline gold plasmonic nanorods.* Optics Express, 2020. **28**(23): p. 34960-34972.

188. Knittel, V., et al., *Nonlinear Photoluminescence Spectrum of Single Gold Nanostructures.* ACS Nano, 2015. **9**(1): p. 894-900.

189. Hensen, M., et al., *Spatial Variations in Femtosecond Field Dynamics within a Plasmonic Nanoresonator Mode.* Nano Letters, 2019. **19**(7): p. 4651-4658.

190. Huber, B., et al., *Space- and time-resolved UV-to-NIR surface spectroscopy and 2D nanoscopy at 1 MHz repetition rate.* Review of Scientific Instruments, 2019. **90**(11).

191. Groß, H., et al., *Near-field strong coupling of single quantum dots.* Science Advances, 2018. **4**(3): p. eaar4906.

192. Feichtner, T., O. Selig, and B. Hecht, *Plasmonic nanoantenna design and fabrication based on evolutionary optimization.* Optics Express, 2017. **25**(10): p. 10828-10842.

193. Ochs, M., et al., *Nanoscale Electrical Excitation of Distinct Modes in Plasmonic Waveguides.* Nano Letters, 2021. **21**(10): p. 4225-4230.

194. René, K., et al. *Directed emission by electrically driven optical antennas*. in *Proc.SPIE*. 2018.

195. Pertsch, P., et al., *Tunable Nanoplasmonic Photodetectors.* Nano Letters, 2022. **22**(17): p. 6982-6987.

196. Ochs, M., et al., *Site-selective functionalization of in-plane nanoelectrode-antennas.* Nanoscale, 2023. **15**(11): p. 5249-5256.

197. Wu, C.-H., et al., *Near-Field Photodetection in Direction Tunable Surface Plasmon Polaritons Waveguides Embedded with Graphene.* 2023. **10**(30): p. 2302707.

198. Grimm, P., et al., *Color-Switchable Subwavelength Organic Light-Emitting Antennas.* Nano Letters, 2022. **22**(3): p. 1032-1038.

199. Zurak, L., et al., *Modulation of surface response in a single plasmonic nanoresonator.* Science Advances, 2024. **10**(36): p. eadn5227.

200. Razinskas, G., et al., *Normal-Incidence PEEM Imaging of Propagating Modes in a Plasmonic Nanocircuit.* Nano Letters, 2016. **16**(11): p. 6832-6837.

201. Dai, W.-H., et al., *Mode Conversion in High-Definition Plasmonic Optical Nanocircuits.* Nano Letters, 2014. **14**(7): p. 3881-3886.

202. Rewitz, C., et al., *Coherent Control of Plasmon Propagation in a Nanocircuit.* Physical Review Applied, 2014. **1**(1): p. 014007.

203. Chen, T.-Y., et al., *Modal Symmetry Controlled Second-Harmonic Generation by Propagating Plasmons.* Nano Letters, 2019. **19**(9): p. 6424-6428.

204. Biagioni, P., et al., *Dynamics of Four-Photon Photoluminescence in Gold Nanoantennas.* Nano Letters, 2012. **12**(6): p. 2941-2947.

205. Wu, X., et al., *On-Chip Single-Plasmon Nanocircuit Driven by a Self-Assembled Quantum Dot.* Nano Letters, 2017. **17**(7): p. 4291-4296.

206. Aeschlimann, M., et al., *Cavity-assisted ultrafast long-range periodic energy transfer between plasmonic nanoantennas.* Light: Science & Applications, 2017. **6**(11): p. e17111-e17111.

207. See, K.-M., F.-C. Lin, and J.-S. Huang, *Design and characterization of a plasmonic Doppler grating for azimuthal angle-resolved surface plasmon resonances.* Nanoscale, 2017. **9**(30): p. 10811-10819.

208. Ouyang, L., et al., *Spatially Resolving the Enhancement Effect in Surface-Enhanced Coherent Anti-Stokes Raman Scattering by Plasmonic Doppler Gratings.* ACS Nano, 2021. **15**(1): p. 809-818.

209. Barman, P., et al., *Nonlinear Optical Signal Generation Mediated by a Plasmonic Azimuthally Chirped Grating.* Nano Letters, 2022. **22**(24): p. 9914-9919.

210. Chakraborty, A., et al., *Broadband Four-Wave Mixing Enhanced by Plasmonic Surface Lattice Resonance and Localized Surface Plasmon Resonance in an Azimuthally Chirped Grating.* Laser & Photonics Reviews, 2023. **17**(7): p. 2200958.

211. Menabde, S.G., et al., *Low-Loss Anisotropic Image Polaritons in van der Waals Crystal α-MoO3.* 2022. **10**(21): p. 2201492.

212. Li, C., S. Bolisetty, and R. Mezzenga, *Hybrid nanocomposites of gold single-crystal platelets and amyloid fibrils with tunable fluorescence, conductivity, and sensing properties.* Advanced Materials, 2013. **25**(27): p. 3694-3700.

213. Chen, Y., et al., *Guiding growth orientation of two-dimensional Au nanocrystals with marine chitin nanofibrils for ultrasensitive and ultrafast sensing hybrids.* Journal of Materials Chemistry B, 2017. **5**(48): p. 9502-9506.

214. Boya, R., D. Jayaraj, and G.U. Kulkarni, *Top-contacting molecular monolayers using single crystalline Au microplate electrodes.* Chemical Science, 2013. **4**(6).

215. Seo, H.J., et al., *Ultrathin silver telluride nanowire films and gold nanosheet electrodes for a flexible resistive switching device.* Nanoscale, 2018. **10**(12).

216.    Radha, B., et al., *Large-area ohmic top contact to vertically grown nanowires using a free-standing Au microplate electrode.* ACS Applied Materials and Interfaces, 2012. **4**(4).

217.    Zhu, Y., et al., *On-chip single-crystal plasmonic optoelectronics for efficient hot carrier collection and photovoltage detection.* Light: Science & Applications, 2025. **14**(1): p. 325.

218.    Kim, J., et al., *Wearable biosensors for healthcare monitoring.* Nature Biotechnology, 2019. **37**(4): p. 389-406.

219.    Nyström, G., et al., *Amyloid Templated Gold Aerogels.* Advanced Materials, 2016. **28**(3): p. 472-478.

220.    Singh, A., M. Chaudhari, and M. Sastry, *Construction of conductive multilayer films of biogenic triangular gold nanoparticles and their application in chemical vapour sensing.* Nanotechnology, 2006. **17**(9).

221.    Schechter, I., M. Ben-Chorin, and A. Kux, *Gas Sensing Properties of Porous Silicon.* Analytical Chemistry, 1995. **67**(20): p. 3727-3732.

222.    Goyal, R.N., A. Aliumar, and M. Oyama, *Comparison of spherical nanogold particles and nanogold plates for the oxidation of dopamine and ascorbic acid.* Journal of Electroanalytical Chemistry, 2009. **631**(1-2).

223.    Momeni, S., et al., *Gold nanosheets synthesized with red marine alga Actinotrichia fragilis as efficient electrocatalysts toward formic acid oxidation.* RSC Advances, 2016. **6**(79): p. 75152-75161.

224.    Wenjing, L., et al., *Fabrication of gold nanoprism thin films and their applications in designing high activity electrocatalysts.* Journal of Physical Chemistry C, 2009. **113**(5): p. 1738-1745.

225.    Li, M., et al., *Single-crystal Au microflakes modulated by amino acids and their sensing and catalytic properties.* Journal of Colloid and Interface Science, 2016. **467**: p. 115-120.

226.    Jeong, W., et al., *Ultraflat Au nanoplates as a new building block for molecular electronics.* Nanotechnology, 2016. **27**(21).

227.    Xia, X., et al., *A silver nanocube on a gold microplate as a well-defined and highly active substrate for SERS detection.* Journal of Materials Chemistry C, 2013. **1**(38): p. 6145-6150.

228.    Singh, A., R. Pasricha, and M. Sastry, *Ultra-low level optical detection of mercuric ions using biogenic gold nanotriangles.* Analyst, 2012. **137**(13): p. 3083-3090.

229.    Lin, W.H., Y.H. Lu, and Y.J. Hsu, *Au nanoplates as robust, recyclable SERS substrates for ultrasensitive chemical sensing.* Journal of Colloid and Interface Science, 2014. **418**.

230. Zhou, M., et al., *Thickness-dependent SERS activities of gold nanosheets controllably synthesized via photochemical reduction in lamellar liquid crystals.* Chemical Communications, 2015. **51**(24): p. 5116-5119.

231. Fang, G., et al., *Formation of different gold nanostructures by silk nanofibrils.* Materials Science and Engineering: C, 2016. **64**: p. 376-382.

232. Xin, W., et al., *Novel Strategy for One-Pot Synthesis of Gold Nanoplates on Carbon Nanotube Sheet As an Effective Flexible SERS Substrate.* ACS Applied Materials and Interfaces, 2017. **9**(7).

233. Chen, Y., et al., *A large-scale, flexible and two-dimensional AuNP/NS as a highly active and homogeneous SERS substrate.* Applied Physics Express, 2019. **12**(7): p. 075005.

234. Hwang, A., et al., *Atomically Flat Au Nanoplate Platforms Enable Ultraspecific Attomolar Detection of Protein Biomarkers.* ACS Applied Materials and Interfaces, 2019. **11**(21).

235. Wang, J., et al., *Mono-metal epitaxial growth for hollow gold microsheets and their hybrids for surface-enhanced Raman scattering.* Applied Physics A, 2020. **126**(6): p. 467.

236. Eom, G., et al., *Ultrasensitive Detection of Ovarian Cancer Biomarker Using Au Nanoplate SERS Immunoassay.* Biochip Journal, 2021. **15**(4): p. 348-355.

237. He, S., et al., *Photochemical strategies for the green synthesis of ultrathin Au nanosheets using photoinduced free radical generation and their catalytic properties.* Nanoscale, 2018. **10**(39): p. 18805-18811.

238. Lin, F.-C., et al., *Designable Spectrometer-Free Index Sensing Using Plasmonic Doppler Gratings.* Analytical Chemistry, 2019. **91**(15): p. 9382-9387.

239. Raman, C.V. and K.S. Krishnan, *A new type of secondary radiation.* Nature, 1928. **121**(3048): p. 501-502.

240. Fleischmann, M., P.J. Hendra, and A.J. McQuillan, *Raman spectra of pyridine adsorbed at a silver electrode.* Chemical Physics Letters, 1974. **26**(2): p. 163-166.

241. Crivianu-Gaita, V. and M. Thompson, *Aptamers, antibody scFv, and antibody Fab' fragments: An overview and comparison of three of the most versatile biosensor biorecognition elements.* Biosensors & Bioelectronics, 2016. **85**: p. 32-45.

242. Dahanayaka, D.H., et al., *Optically transparent Au{111} substrates: Flat gold nanoparticle platforms for high-resolution scanning tunneling microscopy.* Journal of the American Chemical Society, 2006. **128**(18).

243. Zhang, L., et al., *Dynamic molecular tunnel junctions based on self-assembled monolayers for high tunneling current triboelectricity generation.* Journal of Materials Chemistry A, 2023. **11**(10).

244. Ren, B., et al., *Tip-enhanced raman spectroscopy of benzenethiol adsorbed on Au and Pt single-crystal surfaces.* Angewandte Chemie - International Edition, 2004. **44**(1).

245.  Deckert-Gaudig, T. and V. Deckert, *Ultraflat transparent gold nanoplates - Ideal substrates for tip-enhanced Raman scattering experiments.* Small, 2009. **5**(4): p. 432-436.

246.  Richard-Lacroix, M. and V. Deckert, *Direct molecular-level near-field plasmon and temperature assessment in a single plasmonic hotspot.* Light: Science and Applications, 2020. **9**(1).

247.  Stöckle, R.M., et al., *Nanoscale chemical analysis by tip-enhanced Raman spectroscopy.* Chemical Physics Letters, 2000. **318**(1): p. 131-136.

248.  Pettinger, B., et al., *Nanoscale probing of adsorbed species by tip-enhanced Raman spectroscopy.* Physical Review Letters, 2004. **92**(9).

249.  Pashaee, F., et al., *Tip-enhanced Raman spectroscopy of self-assembled thiolated monolayers on flat gold nanoplates using gaussian-transverse and radially polarized excitations.* Journal of Physical Chemistry C, 2013. **117**(30).

250.  Pashaee, F., et al., *Tip-enhanced Raman spectroscopy of graphene-like and graphitic platelets on ultraflat gold nanoplates.* Physical Chemistry Chemical Physics, 2015. **17**(33).

251.  Pienpinijtham, P., et al., *Micrometer-sized gold nanoplates: Starch-mediated photochemical reduction synthesis and possibility of application to tip-enhanced Raman scattering (TERS).* Physical Chemistry Chemical Physics, 2012. **14**(27): p. 9636-9641.

252.  Haruta, M., et al., *NOVEL GOLD CATALYSTS FOR THE OXIDATION OF CARBON-MONOXIDE AT A TEMPERATURE FAR BELOW 0-DEGREES-C.* CHEMISTRY LETTERS, 1987(2): p. 405-408.

253.  Sanchez, A., et al., *When Gold Is Not Noble: Nanoscale Gold Catalysts.* The Journal of Physical Chemistry A, 1999. **103**(48): p. 9573-9578.

254.  Hashmi, A.S.K. and G.J. Hutchings, *Gold catalysis.* ANGEWANDTE CHEMIE-INTERNATIONAL EDITION, 2006. **45**(47): p. 7896-7936.

255.  Primo, A., et al., *One-Step Pyrolysis Preparation of 1.1.1 Oriented Gold Nanoplatelets Supported on Graphene and Six Orders of Magnitude Enhancement of the Resulting Catalytic Activity.* Angewandte Chemie - International Edition, 2016. **55**(2).

256.  Kanuru, V.K., et al., *Sonogashira Coupling on an Extended Gold Surface in Vacuo: Reaction of Phenylacetylene with Iodobenzene on Au(111).* Journal of the American Chemical Society, 2010. **132**(23): p. 8081-8086.

257.  Bhosale, M.A., D.R. Chenna, and B.M. Bhanage, *One–step sonochemical irradiation dependent shape controlled crystal growth study of gold nano/microplates with high catalytic activity in degradation of dyes.* ChemistrySelect, 2016. **1**(3).

258.  Zhu, J., et al., *Gold Flake-Enabled Miniature Capacitive Picobalances.* Small Methods, 2024.

259.   Singh, A.V., et al., *Biomineralized anisotropic gold microplate-macrophage interactions reveal frustrated phagocytosis-like phenomenon: A novel paclitaxel drug delivery vehicle.* ACS Applied Materials and Interfaces, 2014. **6**(16).

260.   Radha, B., et al., *Movable Au microplates as fluorescence enhancing substrates for live cells.* Nano Research, 2010. **3**(10).

261.   Jucker, L., et al., *Development of Rotaxanes as E-Field-Sensitive Superstructures in Plasmonic Nano-Antennas.* Organic Materials, 2022. **4**(03): p. 127-136.

262.   Ah, C.S., et al., *Size-controlled synthesis of machinable single crystalline gold nanoplates.* Chemistry of Materials, 2005. **17**(22).

263.   Yun, Y.J., et al., *Fabrication of versatile nanocomponents using single-crystalline Au nanoplates.* Applied Physics Letters, 2005. **87**(23).

264.   Joo, J.H., et al., *Aqueous epitaxial growth of ZNO on single crystalline AU microplates.* Crystal Growth and Design, 2013. **13**(3).

265.   Sharma, V., et al., *Large-scale optothermal assembly of colloids mediated by a gold microplate.* Journal of Physics Condensed Matter, 2020. **32**(32).

266.   Wu, X., et al., *Light-driven microdrones.* Nature Nanotechnology, 2022. **17**(5): p. 477-484.

267.   Qin, J., et al., *Light-driven plasmonic microrobot for nanoparticle manipulation.* Nature Communications, 2025. **16**(1): p. 2570.

268.   Nootchanat, S., et al., *Formation of large H2O2-reduced gold nanosheets via starch-induced two-dimensional oriented attachment.* RSC Advances, 2013. **3**(11).

269.   Capitaine, A. and B. Sciacca, *Nanocube Epitaxy for the Realization of Printable Monocrystalline Nanophotonic Surfaces.* Advanced Materials, 2022. **34**(24): p. 2200364.

270.   V. Grayli, S., et al., *Scalable, Green Fabrication of Single-Crystal Noble Metal Films and Nanostructures for Low-Loss Nanotechnology Applications.* ACS Nano, 2020. **14**(6).

271.   Neal, R.D., et al., *Large-Area Periodic Arrays of Atomically Flat Single-Crystal Gold Nanotriangles Formed Directly on Substrate Surfaces.* Small, 2022. **18**(52): p. 2205780.

272.   Lawson, Z.R., et al., *Light-Mediated Growth of Gold Nanoplates Carried Out in Total Darkness.* ACS Nano, 2025.

273.   Golze, S.D., et al., *Plasmon-Mediated Synthesis of Periodic Arrays of Gold Nanoplates Using Substrate-Immobilized Seeds Lined with Planar Defects.* Nano Letters, 2019. **19**(8): p. 5653-5660.

274. Lawson, Z.R., et al., *Large-area arrays of epitaxially aligned silver nanotriangles seeded by gold nanostructures.* Materials Chemistry Frontiers, 2024. **8**(9): p. 2149-2160.

275. Chen, Y.Q., et al., *Sub-10 nm fabrication: methods and applications.* International Journal of Extreme Manufacturing, 2021. **3**(3).

276. Ye, X., et al., *Morphologically controlled synthesis of colloidal upconversion nanophosphors and their shape-directed self-assembly.* Proceedings of the National Academy of Sciences of the United States of America, 2010. **107**(52).

277. Kim, J., et al., *Polymorphic Assembly from Beveled Gold Triangular Nanoprisms.* Nano Letters, 2017. **17**(5): p. 3270-3275.

278. Zhou, Y., et al., *Shape-selective deposition and assembly of anisotropic nanoparticles.* Nano Letters, 2014. **14**(4).

279. Zhan, P.F., et al., *DNA Origami Directed Assembly of Gold Bowtie Nanoantennas for Single-Molecule Surface-Enhanced Raman Scattering.* Angewandte Chemie-International Edition, 2018. **57**(11): p. 2846-2850.

280. Morales, M.A. and J.M. Halpern, *Guide to Selecting a Biorecognition Element for Biosensors.* Bioconjugate Chemistry, 2018. **29**(10): p. 3231-3239.